\documentstyle[11pt,aaspp]{article}

\begin{document}

\hyphenation{PSPC}
\hyphenation{X-ray}

\slugcomment{To Appear in MNRAS}

\title{Cosmological Implications of Galaxy Cluster Evolution}

\author{John C. Tsai}
\affil{Canadian Institute for Theoretical Astrophysics}
\affil{McLennan Labs, University of Toronto, 60 St. George St.}
\affil{Toronto, Ontario, Canada, M5S 1A7}
\affil{tsai@cita.utoronto.ca}
%\affil{Canadian Institute for Theoretical Astrophysics}
%\authoraddr{McLennan Labs, University of Toronto, 60 St. George St.,
%Toronto, Ontario, Canada, M5S 1A7, tsai@cita.utoronto.ca}
\vskip 6pt

\author{David A. Buote\altaffilmark{1}}
%\affil{Massachusetts Institute of Technology}
\affil{Department of Physics and Center for Space Research 37-241}
\affil{Massachusetts Institute of Technology}
\affil{77 Massachusetts Avenue, Cambridge, MA 02139}
\affil{dbuote@space.mit.edu}
%\authoraddr{Department of Physics
%and Center for Space Research 37-241, Massachusetts Institute of
%Technology, 77 Massachusetts Avenue, Cambridge, MA 02139, 
%dbuote@space.mit.edu}

\altaffiltext{1}{Present Address: Institute of Astronomy, Madingly
Road, Cambridge CB3 0HA, UK}

\begin{abstract}

We analyze with hydrodynamical simulations the evolution of galaxy
clusters in a cosmological environment.  Power ratios (Buote \& Tsai
1995) are used to quantitatively relate cluster morphologies to their
dynamical states. The simulated clusters follow the same
``evolutionary track'' obeyed by a sample of low-redshift $(z<0.2)$
ROSAT PSPC clusters (Buote \& Tsai 1996) indicating that the detailed
evolution of individual simulated clusters is consistent with observed
clusters.  However, the distribution of simulated clusters (for
$\Omega=1$ standard Cold Dark Matter) along the evolutionary track at
the present epoch, which indicates a measure of the present balance of
cluster formation and relaxation rates, suggests that there are too
many simulated clusters with significant amounts of substructure to be
consistent with the observations, thus favoring a lower value of
$\Omega$.  Perpendicular to the evolutionary track the distributions
of observed and simulated clusters are consistent which may indicate a
success of the cosmological model (e.g., power spectrum). Analysis of
high-redshift simulated clusters suggests that the distribution of
clusters both along and perpendicular to the evolutionary track is
effectively constant from $z\sim 0.6$ to the present but changes
significantly for $z > 0.6$.
\end{abstract}


%\vfill\eject

\section{Introduction\label{Intro}}

In a universe where structure formation occurs by hierarchical 
clustering, e.g. in the standard cold dark matter cosmogony 
(CDM), newly formed galaxy clusters should have complex morphologies 
with copious amounts of small-scale structure (i.e. ``substructure'').  
If the cluster relaxation time is short ($\la$ Gyr), which is reasonable 
since the sound crossing and dynamical times are short, then
the fraction of clusters which appear unrelaxed at a given epoch will
reflect the cluster formation rate at that time.  This rate
likely depends upon the cosmological density parameter $\Omega$
(e.g. Richstone, Loeb, \& Turner 1992; Lacey \& Cole 1993; Bartelmann,
Ehlers, \& Schneider 1993; but also see Kauffmann \& White 1993; 
Nakamura, Hattori, \& Mineshige 1995), thus the distribution of 
cluster morphologies provides a constraint on the density of the universe.

Alternatively, if structure formation proceeds by some other means, such
as the fragmentation (top--down) picture of hot dark matter,
cosmic strings, or textures, the morphology of clusters may
also provide cosmological constraints.  In the top--down scenario, 
for example, structures which are on the verge of fragmenting into 
smaller components should exhibit significant substructure.  Hence
the fraction of unrelaxed structures measures the ``destruction'' 
rate of these objects and this rate may relate to the underlying 
cosmology.

Implementing the above ideas requires a suitable measure of
substructure which relates directly to the dynamical state of the
cluster. The degree of virialization is well defined by the shape of
the cluster potential: i.e. a spherically symmetric or moderately
flattened potential implies a relaxed cluster whereas very distorted
isopotential contours imply an unvirialized state (e.g., Buote \& 
Tsai 1995 and Buote \& Canizares 1996). The potential is characterized by
a multipole expansion.  The size of the
lowest few terms of the expansion relative to the monopole term gives
an indication of the distortions in the potential; very small scale
structure represented by higher order terms in the expansion
contributes negligibly to the overall cluster potential.  Since three
dimensional properties of individual clusters cannot be determined
from observations, attention must be limited to the projected cluster
potential and to its multipoles.  The corresponding loss of
information due to projection effects can only be recovered by
considering a statistically large sample of clusters.

A measure of substructure based on expansions of the potential was 
introduced by Buote \& Tsai (1995a; hereafter BTa).  The 
two dimensional potential $\Psi(R,\phi)$ due to material interior to
projected radius $R$ is expanded,
\begin{equation}
\Psi(R,\phi) = -2Ga_0\ln\left({1 \over R}\right) -2G
\sum^{\infty}_{m=1} {1\over m R^m}\left(a_m\cos m\phi + b_m\sin
m\phi\right), \label{eqn.multipole}
\end{equation}
where $\phi$ is the azimuthal angle, $G$ is the gravitational constant
and,
\begin{eqnarray}
a_m(R) & = & \int_{R^{\prime}\le R} \Sigma(\vec x^{\prime})
\left(R^{\prime}\right)^m \cos m\phi^{\prime} d^2x^{\prime}\label{eqn.mom1},\\
b_m(R) & = & \int_{R^{\prime}\le R} \Sigma(\vec x^{\prime})
\left(R^{\prime}\right)^m \sin m\phi^{\prime} d^2x^{\prime}\label{eqn.mom2}.
\end{eqnarray}
In the above, $\Sigma$ is the projected mass density.  The square of
each term on the right hand side of eq.  (\ref{eqn.multipole})
integrated over the boundary of a circular aperture of radius
$R_{ap}$ is given by,
\begin{equation}
P_m={1\over 2m^2 R^{2m}_{ap}}\left( a^2_m + b^2_m\right)\label{eqn.powerm}
\end{equation}
for $m>0$ and
\begin{equation}
P_0=\left[a_0\ln\left(R_{ap}\right)\right]^2\label{eqn.power0}
\end{equation}
for $m=0$.  The importance of the term of order $m$ relative to the
monopole term is given by the ratio $P_m/P_0$ (dubbed a ``power ratio'').

The above formalism implies that $\Psi$ is specified by
$\Sigma(R,\phi)$, which can be determined in principle from weak
lensing measurements (e.g., Kaiser \& Squires 1993).  However,
$\Sigma(R,\phi)$ is only available for $\sim 5-10$ clusters and is
given over relatively small regions of the cluster ($\sim 0.4 h_{50}^{-1}$
Mpc; Squires et al. 1995 and references therein).  A more readily
available tracer of the distribution of matter must instead be used
for statistical analysis.  A possibility is the projected density of
galaxies in the cluster.  Although previously used to study cluster
structure (e.g., Geller \& Beers 1982; Dressler \& Shectman 1988; West
\& Bothun 1990), the method is fundamentally limited by the finite
number of galaxies ($N\sim 1000$) in each cluster.  Alternatively, the
X--ray surface brightness profile $\Sigma_x$ traces the projected
X--ray emission measure.  For instruments such as the ROSAT Position
Sensitive Proportional Counter (PSPC; Pfeffermann 1987), this
translates into the projected square of the gas density $\rho^2$ since
the emissivity in the specified observing band for clusters is nearly
constant.  Replacing $\Sigma(R,\phi)$ with $\Sigma_x$ in
eqs. (\ref{eqn.mom1}) and (\ref{eqn.mom2}) implies that the expansion
in eq.  (\ref{eqn.multipole}) will not give precisely the potential
$\Psi$, but some related quantity which would still specify the
dynamical state of the cluster provided the X--ray emission
qualitatively traces the mass distribution (BTa and see below for
explicit tests of this relation).  The abundance of high quality X-ray
imaging data over large regions of observed clusters, especially from
ROSAT, currently makes $\Sigma_x$ the most attractive tracer of
cluster mass.

Buote \& Tsai (1996; hereafter BTb) computed power ratios for a sample
of 59 clusters observed by the ROSAT PSPC.  The primary purpose of
this study was to construct a large data set suitable for
statistical studies of cosmology.  However, the data also exhibited
several interesting properties and correlations which
may have significant implications for cluster evolution.  The
expansion coefficients $a_m$ and $b_m$ were computed on fixed
apertures of $R_{ap}=0.5$, $1.0$, and $1.5$ Mpc (we assume
$H_0 = 80~\rm km~ s^{-1}~Mpc^{-1}$ for the remainder of this paper)
with the center of the aperture defined by, (i) the centroid of the
image (where $P_1=0$), and (ii) the emission peak.  These latter power
ratios are denoted $P_m^{(pk)}$ to distinguish them from those of case
(i). We compute the power ratios in fixed apertures to enable
consistent comparison of structures of the given scale.  This
procedure is a key aspect of the power ratios: the scale of
substructures being probed is set by the aperture size so that
different information is generally provided by power ratios computed
in apertures of different sizes (BTa).

The set of ratios $P_m/P_0$ (with 1 Mpc apertures defined about the
centroid) for $m=2$, 3, and 4 define the axes of a three dimensional
space in which each cluster occupies a unique position (we actually
use $log(P_m/P_0)$ for the axes).  Those clusters with significant
non--axisymmetric structure will generally have larger values of
$P_m/P_0$ so will occupy the region far removed from the origin of the
coordinate system.  Conversely, relaxed clusters will congregate
around the origin.  BTb found that observed clusters of varying
morphologies occupied only a restricted region in the space defined by
the power ratios, lying basically in a thick straight filament
extending outwards from the origin.  The projection of this filament
onto the three coordinate planes implies significant correlations
among the three power ratios.  In particular, the correlation between
$P_2/P_0$ and $P_4/P_0$ is rather strong and admits a physical
interpretation.  Recently formed clusters arrive on the correlation
line with large values for the power ratios.  As structure is erased
by relaxation, the cluster moves down the correlation line towards the
origin, finally coming to rest near the origin as a virialized
cluster.  Of course, clusters can also move upward toward large power
ratios if, for example, the cluster accretes or merges with a
companion previously outside the aperture being considered.  The
correlation line (or indeed the thick filament in the three
dimensional space) is interpreted as the ``evolutionary track''
followed by clusters as they relax and age.  The position along this
track specifies a measure of the dynamical state of the cluster and
the distribution of clusters along the track provides a measure of the
balance of the formation and relaxation rates for the clusters (see \S
\ref{SClow1}). This distribution should be sensitive to $\Omega$ (see
above).

In this paper, we use N--body/Hydrodynamical simulations of cluster
formation in the $\Omega=1$ CDM cosmogony (Navarro, Frenk, \& White
1995; hereafter NFW) to (i) understand the correlations of the
observed power ratios and test the evolutionary track interpretation
of the correlation line, (ii) determine if clusters formed in an
$\Omega=1$ CDM cosmogony are similar in structure to real clusters,
(iii) to assess the cosmological implications of the observed
distribution of power ratios, and (iv) to study the the evolution of
power ratios with redshift in an $\Omega=1$ CDM cosmogony.  The
cosmological consequences of substructure have been previously
considered by several groups (e.g., Evrard et al. 1993, Mohr et
al. 1995; Jing et al. 1995) using a different set of statistics from
the power ratios (e.g., centroid shifts, axial ratios, orientation
angles, and radial falloff).  While Mohr et al. indicate that the
$\Omega=1$ CDM model is preferred over low $\Omega$ models, Jing et
al. reach the different conclusion that an $\Omega=0.3$, flat model
appears to work equally well as the $\Omega=1$ case.  Although similar
statistics are used, the reasons for the discrepancy are not clear.
Some differences are that the apertures over which the statistics are
computed are not treated the same way and Jing et al.  use a
hydrostatic model for putting gas into purely dissipationless
simulations, whereas Mohr et al. use gas-dynamical simulations.
Clearly, an independent set of statistics should be applied to the
problem to check previous results.  This is also important, however,
because the previously used statistics are not clearly related to the
dynamical state of the cluster (see BTa), although they do represent a
means of quantitatively classifying the structures present.  For
example, it is not obvious that centroid shifts and axial ratios admit
an interpretation in terms of the evolutionary track discussed above.
The X--ray data used ({\it Einstein} IPC images) also have inferior
spatial resolution and signal-to-noise ratios to the ROSAT data
considered in BTb.

In \S \ref{method}, we briefly discuss the simulations and our method
of analysis.  The results for an aperture of 1 Mpc are presented in \S
\ref{onempc}.  The power ratios computed on the dark
matter distribution are compared with those for the gas in 
\S\ref{dark.gas}.  A discussion and conclusions are given in \S
\ref{disscon}

\section{Method \label{method}}

The simulations are a combined N--body/SPH calculation of 6 clusters
with velocity dispersions spanning the typically observed range,
$v\sim 400 - 1300~\rm km~s^{-1}$.  The initial conditions for the
cluster simulation were selected from a large N--body simulation
without bias regarding the morphology of the cluster at the present
time (specified by $\sigma_8=0.63$).  The baryonic fraction is assumed
to be $\Omega_b=0.1$.  Cooling is not included in the simulations so
structures in the densest regions (cluster cores) are not well
modeled.  This is, however, not a problem since we never consider
scales much smaller than $r\sim 100 ~\rm kpc$ which is in any event
roughly the gravitational softening length of the
simulations.  See NFW for a detailed description of the simulations.

We simulate observations of each cluster along three orthogonal axes
with random orientation.  The X--ray surface brightness is computed,
\begin{equation}
\Sigma_x(R,\phi)=\int_{-\infty}^{\infty}\Lambda_x \left[T\left(R, \phi,
z\right)\right] \rho^2 dz,
\end{equation}
where $\rho$ is the gas density, $R$ and $\phi$ specify the location
in the observed plane, and $z$ is along the line of sight.  The
quantity $\Lambda_x$ is the cluster emissivity convolved with the
spectral response of the PSPC.  Although the atomic X--ray emissivity
of the gas generally varies with gas temperature, the PSPC convolved
emissivity is constant to a very good approximation for the clusters
considered here (most cluster gas is hotter than $\sim 1$ keV where
the PSPC is most sensitive; see NRA 91-OSSA-3, Appendix F, ROSAT
mission description).  We take $\Lambda_x$ to be strictly constant,
and since we consider exclusively ratios of $P_m$, the exact numerical
value of $\Lambda_x$ is immaterial.  Reasonable amounts of Poisson
noise and undetected point sources have no effect on the computed
$P_m$ (see BTa) so are not included in constructing the synthetic
observations.

The power ratios are computed from $\Sigma_x$ for each of the
projections and for a variety of redshifts $z$ in an aperture of
radius 1.0 Mpc in order to test the ``evolutionary track'' idea.
Although we also compute the power ratios for an aperture size of 0.5
Mpc, these results are in every way qualitatively similar to those
with the larger aperture so are not presented in this paper.
Furthermore, there are too few observed clusters with well determined
$P_m/P_0$ for an aperture size of 1.5 Mpc to be considered.  We note
that other choices for $R_{ap}$ are possible: for example, the virial
radius as fixed by the temperature of the X--ray emitting gas.  As
noted in BTa, however, there are significant difficulties with this
latter approach.  Uncertainties in the temperature imply large
uncertainties in the determined virial radius.  This potentially masks
the information contained in the distributions of power ratios.
Secondly, typical virial radii are large enough that in many cases
there is insufficient available data to accurately compute the power
ratios.  We therefore restrict $R_{ap}$ to fixed metric distances.

\section{Results for a 1 Mpc Aperture\label{onempc}}

\subsection{Cluster Evolution and the Evolutionary Track\label{CL1}}

In this section we describe the morphological evolution of
the NFW clusters to examine how the dynamical states of the clusters
are indicated in terms of the power ratios, and to test
the ``evolutionary track'' interpretation of BTb.  The evolution of the
largest simulated cluster (cluster CL1 in the notation of NFW) is
shown in Figures \ref{clus1x}, \ref{clus1y}, and \ref{clus1z}.  The
SPH gas particles are plotted as viewed along the three orthogonal
directions $x$, $y$, and $z$.  The evolution of the cluster is typical
for bottom--up structure formation scenarios.  At early times, the
cluster is a loose aggregate of small clumps which are growing
constantly by accreting surrounding gas (and dark matter).  At
$z=0.49$ two of the clumps merge followed by the final merger
event somewhere between $z=0.49$ and $z=0.35$.  The
subsequent evolution of the cluster is simply that of a single clump
achieving hydrostatic equilibrium while incrementally growing by the
slow accretion of adjacent material.  We only show this cluster in
detail since the evolution and growth of the other clusters proceeds
basically in the same manner as described above, although the
morphology at $z=0$ is generally very different from that of CL1
(see NFW).

Some of the power ratios computed for this cluster for an aperture
size of $R_{ap}=1~\rm Mpc$ are shown in Figures \ref{powrat42} and
\ref{powrat32}.  The heavy solid lines indicate the limited region of
$P_m$ space occupied by observed clusters (hence the observed
correlation among the $P_m$).  These are simply made by joining the
ends of several of the error bars from Figure 4 of BTb; it is only
meant to give a rough idea of the observed values of $P_m$.  We
consider each projection in turn to understand variations in the $P_m$
due to both cluster evolution and differing viewing directions.

\subsubsection{Power Ratios for Three Different Projections}

At the earliest redshift $z=1.4$, the cluster viewed along the
$x$ axis has widely separated clumps.  We define clumps as belonging
to the same cluster if they satisfy the following condition.  The 
surface brightness profile $\Sigma_x$ of each clump is circularly 
averaged about the local peak of $\Sigma_x$ to yield a radial profile.  
If any point on the annulus which has $\Sigma_x$ equal to 1\% of the 
peak value is within 1 Mpc of the emission peak of another clump, then 
those two clumps are considered part of the same cluster.  For two 
given clumps, if the previous condition is valid for annuli of either 
clump, they are considered part of the same cluster.  Although more
arbitrary than some dynamically motivated condition, this definition
allows a consistent comparison of theory to observations provided the
latter are subject to the same criteria.  This definition also does
not require perhaps dubious dynamical modeling of the cluster.

The 3 major clumps at $z=1.4$ are considered separate clusters
by the above definition, although after subsequent mergers the
distinction disappears.  The power ratios shown in the figures are
computed for the central clump.  The morphology is ``boxy'' or
elongated along the four filaments extending from the clump and there
is considerable small scale structure compared to the overall
ellipticity of the clump.  This is reasonable considering the
clump recently formed out of diffuse matter and is not yet virialized.
Corresponding to these properties, the values of $P_4/P_0$ and
$P_3/P_0$ are above average for the amount of $P_2/P_0$ present and
the clump sits at the upper boundary of the region occupied by
observed clusters.  Although the $P_2/P_0$ value is reminiscent of a
cluster like A545 (see BTb) which is highly elongated, there is
considerably more higher order structure than in A545.  By the next
redshift $z=0.95$, the central clump has relaxed and the higher
order non--axisymmetric structures are erased.  All the ratios
decrease and the cluster moves into the region of $P_m$ space occupied
by virialized, regular clusters such as A2029.  At $z=0.68$, the
three clumps are still considered separate, however, because they have
moved closer together the structures linking the clumps move into the
aperture and contribute to an increase in the power ratios.

Shortly after $z=0.68$, one of the other clumps moves into the
aperture of the central clump and the two gradually merge.  The next
redshift considered ($z\sim 0.49$) shows the final stages of
this merger.  The cluster morphology implies a large quadrupole moment
$P_2/P_0$, but not a large value of $P_3/P_0$.  The latter result
occurs because the two clumps in the merger are nearly equal-sized so
there is little structure that is symmetric with respect to rotations
of angle $2\pi/3$ about the origin -- $P_3/P_0=0$ for bimodals with
exactly equal-sized clumps (BTa).  Note that the evolution of the
power ratios in time does not necessarily follow the dashed lines in
Figures \ref{powrat42} and \ref{powrat32}; these lines simply
emphasize the sequence of circles which capture the cluster structure
at a fixed set of redshifts.  For example, the entire merger sequence
has been missed because of the discretely spaced coverage of redshifts.
In actuality, the power ratios must become very large and move to the
upper right of the power ratio plots when the merging cluster is a
bimodal with widely separated clumps.  As the clumps coalesce, the
power ratios move back down towards the lower left.  The motion of the
cluster in power ratio space is thus more complex than indicated
simply by the dotted lines, although we know that the merger time is
short because the merger happens between two closely spaced redshifts.
The above merging sequence will be clearly seen below for the $z$
projection.

\begin{figure}
\caption{  \label{clus1x} }

\plotone{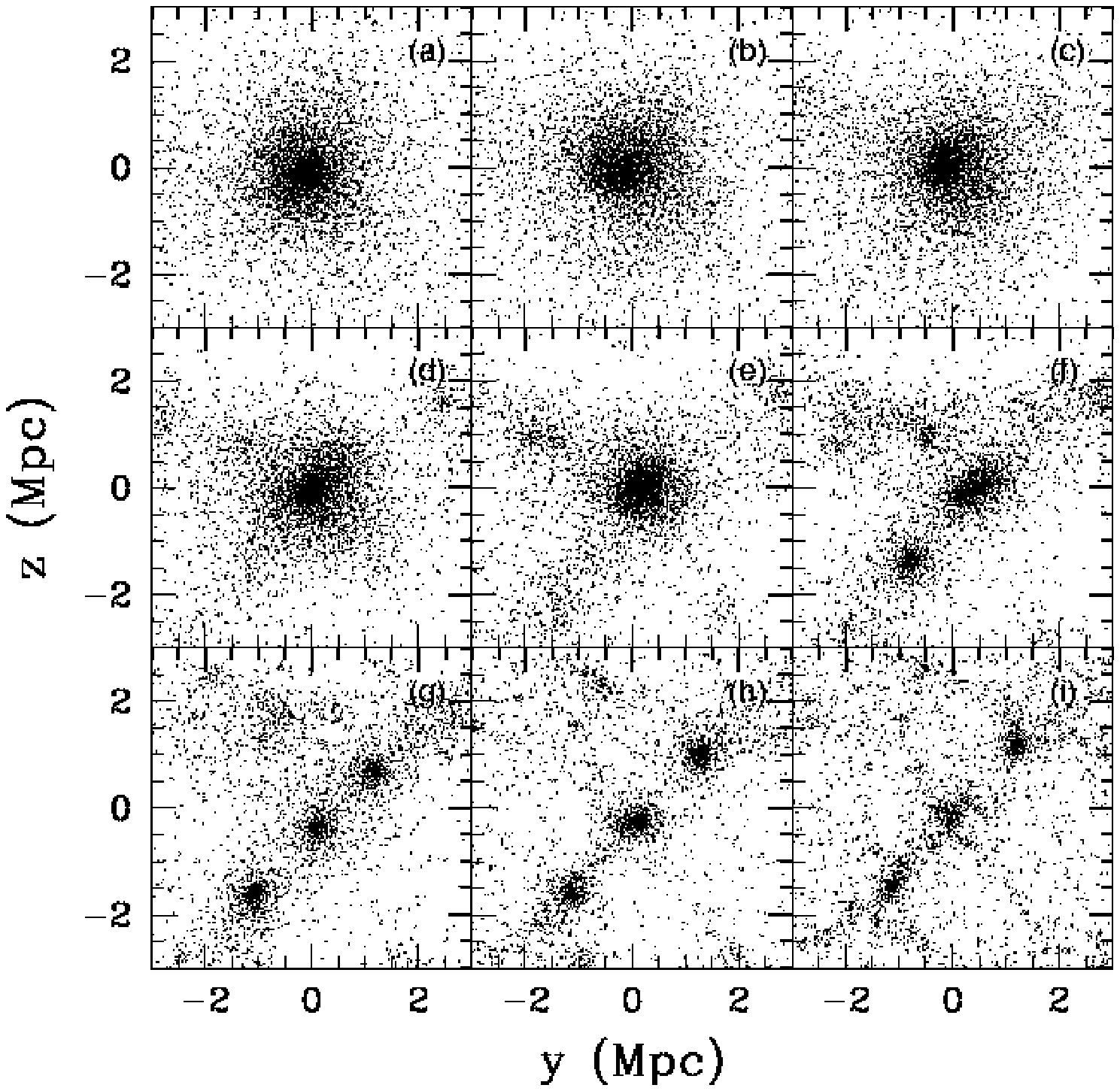}

\raggedright
The SPH gas particles in the largest simulated cluster of Navarro et al. 
(1995; cluster CL1) are shown when viewed down the $x$ axis.  
Distances are given
in physical units of Mpc.  Each panel corresponds to a different redshift:
(a) gives $z = 0$ (the present), (b) gives $z = 0.065$,
(c) gives $z = 0.14$, (d) gives $z = 0.23$, (e) gives 
$z = 0.35$, (f) gives $z = 0.49$, (g) gives $z = 0.68$,
(h) gives $z = 0.95$, and (i) gives $z = 1.4$.
\end{figure}

\begin{figure}
\caption{  \label{clus1y} }

\plotone{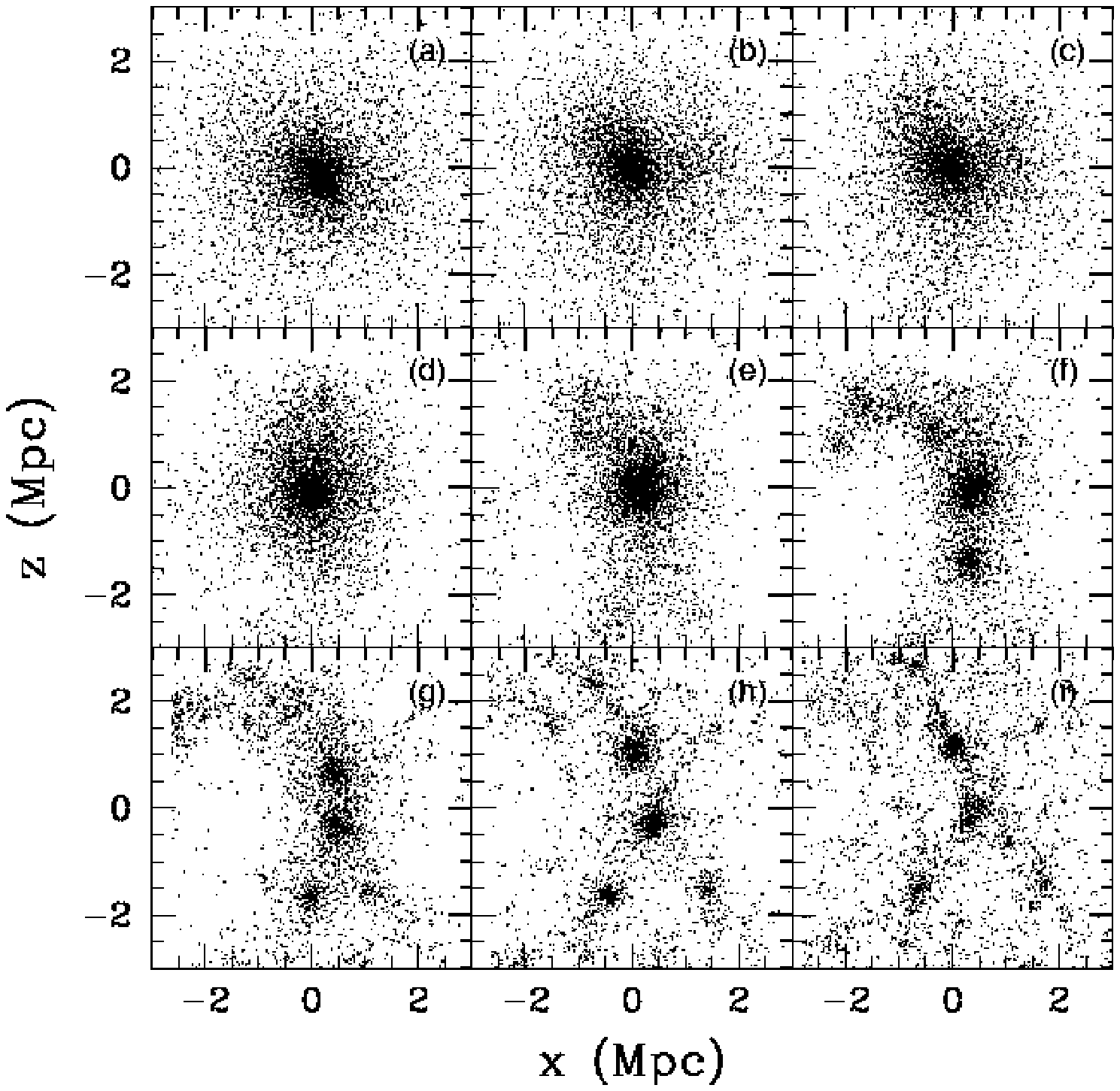}

\raggedright
The SPH gas particles for cluster CL1 are shown viewed down the $y$ axis
for the same set of redshifts as Figure \ref{clus1x}.
\end{figure}

The merger with the last subclump has occurred by $z=0.35$.
Although the cluster shows remaining morphological signs of the event,
it has already relaxed appreciably as is indicated by the smaller
power ratios.  At $z=0.23$ the power ratios grow again because
of the accretion of small surrounding lumps, for example, the gas in
the upper left corner in panel (e) of Figure \ref{clus1x}.  The
remaining relaxation and continued slow accretion of surrounding
material until $z=0$ causes the power ratios to change somewhat
but they remain in the region occupied by relaxed clusters.

The same considerations can be applied to the other projections.
Several forming clumps are again seen at the earliest redshift when
the cluster is viewed down the $y$ projection.  The clumps are
separate according to our criterion (see above) so
for now we plot power ratios of the central clump.  This clump is
unvirialized as it exhibits substantial small-scale structure.
Correspondingly, the values of the higher order power ratios are large
in proportion to the value of $P_2/P_0$ and the clump sits above the
upper boundary of the region occupied by observed clusters in Figures
\ref{powrat42} and \ref{powrat32}.  The clump quickly relaxes so that
by redshift $z=0.95$ the power ratios correspond to that of a
regular cluster.  By the next redshift, the upper clump has moved
sufficiently within the aperture of the central clump so that
the power ratios become large and the cluster moves to the upper right
in the $P_m/P_0$ plots.  The ``cluster'' now consists of both the
central clump and the upper clump which were initially considered
distinct.  This contrasts with cluster properties when viewed
down the $x$ axis.  In that case, the clumps were still sufficiently
far apart to be considered separate clusters hence the power ratios
were computed for only the central clump.

The relaxation and merger of the bimodal cluster at $z=0.68$ has
occurred by the subsequent redshift.  From the $x$ projection, we know
that although the merger is almost complete, two subclumps are clearly
visible (panel [f] of Figure \ref{clus1x}).  However, the two lumps
are along the line of sight when viewed in the $y$ direction, hence
the power ratios are smaller than in the $x$ projection.  The merger
with the remaining secondary lump is complete by $z=0.35$ so
this final merging sequence is not captured in the plots of $P_m/P_0$.
Furthermore, the cluster at $z=0.23$ appears more relaxed in the
$y$ direction than in the $x$ direction so it has correspondingly
smaller power ratios.  The evolution from this time to $z=0$ is
again the virialization of the final cluster.

At the earliest redshift in the $z$ projection, the topmost and 
central clumps are sufficiently close
to be considered a single cluster on which we focus attention.
This is a bimodal cluster with widely separated components of
approximately equal size.  The evolution to $z=0.49$ is simply
the relaxation of this bimodal cluster as the two clumps draw closer
together and merge. At $z=0.49$, the last clump also falls into
the aperture and the cluster again becomes a bimodal but with a
smaller secondary clump.  The cluster evolves in $P_m/P_0$ space by
sliding down the correlation line in either the $P_4/P_0$ - $P_2/P_0$
plane or the $P_3/P_0$ - $P_2/P_0$ plane, confirming the
interpretation of an ``evolutionary track'' followed by clusters.
After this time, the simulated cluster remains in the region occupied
by single-component clusters as continues to evolve quasi-statically.

\subsubsection{Physical Interpretation of the Power Ratios}

We interpret physically the above detailed descriptions of cluster
evolution in the CDM cosmogony as quantified by the power ratios.
Unrelaxed single component clusters, such as the simulated cluster at
$z\sim 1.4$, have significant amounts of smaller scale structure,
which manifests itself as large values of the higher order power
ratios $P_4/P_0$ and $P_3/P_0$ relative to $P_2/P_0$.  These clusters
appear in the power ratio plots in the upper region or above the area
occupied by most observed clusters.  As they virialize, they drop down
to the region occupied by relaxed clusters at the lower left hand
corner of the $P_m/P_0$ plots.  If the clump merges with another
relatively massive clump, the cluster then moves to the upper right
part of the $P_m/P_0$ plots regardless of the individual states of
relaxation of the clumps.  As the clumps merge, the cluster slides
down the correlation line (which has a finite thickness; more about
this below) towards the lower left where the power ratios are small.
Multiple mergers during the lifetime of a cluster are certainly
possible and likely in the CDM cosmogony.  Thus the clusters can move
up and down the correlation line many times in the course of their
lives.  These properties allow us to assess the evolutionary state (in
projection) of any given cluster based on its location in power ratio
space.  For example, unrelaxed single clusters invariably appear in
the upper part of the correlation line.  Examples of this may include
A665 and Cygnus-A (see BTb).

The evolution of any individual cluster can appear differently
depending on viewing direction.  Clumps that are truly widely
separated ($r_{sep}\ga 3~\rm Mpc$) may either appear separated enough
to be considered two distinct clusters, or appear closely
associated and taken to be a single cluster.  Similarly, the merger
rate for a given cluster can significantly differ if in one projection
the secondary clump is outside the aperture whereas in another it is
within the aperture. Thus, in any comparison of simulations to
observations a statistically large number of clusters must be averaged
over to obviate projection effects.  In any case, it is
important to select clusters from observations and simulations in a
consistent fashion to ensure that each sample is similarly affected
by projection effects.

The interpretation of the correlation line in power-ratio space as an
evolutionary track, however, applies regardless of the orientation of
the cluster.  It is simply a quantitative representation of the
evolution of the cluster when viewed in the given direction.  (Note
that projection effectively smoothes the intrinsic three-dimensional
mass distribution of a cluster, hence the power ratios in a given
projection provide a lower limit to the true dynamical state.  A
cluster is no more virialized than indicated by its power ratios.)  We
stress that one of the great assets of the power ratios is the ability
to consistently compare the structures and evolutionary states of
clusters of very different sizes.  For example, the clumps appearing
at $z=1.4$ in the simulation are much less massive than are the
final clusters, however, the power ratios allow for a consistent
relative comparison of the two clusters because the overall surface
brightness does not enter into the ratios $P_m/P_0$.

An intuitive understanding of the evolutionary track may be obtained 
by considering simple analytic models for galaxy
clusters. Figure \ref{powrat42.fit} shows the $P_4/P_0$--$P_2/P_0$
plane along with the region occupied by observed clusters.  We plot
the power ratios of single component $\beta$ models (e.g., Cavaliere
\& Fusco--Femiano 1976),
\begin{equation}
\Sigma_x\left(R\right)\propto \left[1 + \left({R\over 
a_{core}}\right)^2\right]^{-3\beta + 1/2},\label{beta.mod}
\end{equation}
where in order to incorporate models with constant ellipticity
$\epsilon_x$ we take $R^2=x^2 + y^2/q^2$ assuming $q$ is a constant
axis ratio.  We expect that an ellipsoidal distribution with high
ellipticity will relax to a more spherically symmetric distribution
(assuming rotation and anisotropic pressure are dynamically
unimportant for the gas).  Assuming standard values of $\beta=0.75$
and $a_{core}=0.3~\rm Mpc$ (e.g., Jones \& Forman 1984), we plot the
power ratios assuming $\epsilon_x =0.6, 0.3, 0.2,~\rm and ~0.1$.
Higher ellipticities correspond to larger power ratios.  The expected
evolutionary behavior of a simple model translates into a path in
power ratio space very similar to the true evolutionary track for
clusters with a slope that is slightly steeper than the track followed
by the detailed simulations.

To understand this behavior we consider single-component $\beta$
models with $\beta=0.75$ but $a_{core}=0.1~\rm Mpc$.  These are
indicated by the filled circles in Figure \ref{powrat42.fit}.  Again,
the evolution from large to small ellipticities translates into 
decreasing power ratios along a line that is steeper than the 
evolutionary track.  The line followed by the model clusters, however,
is displaced relative to larger core radius case.  A bimodal 
cluster model composed of two spherical $\beta$ model components 
each with $\beta=0.75$ and $a_{core}=0.3~\rm Mpc$ separated by 
distance $r_{sep}$ evolves along a similar track as $r_{sep}$ 
decreases (the open stars of Figure \ref{powrat42.fit}).  Although
each of the preceding models exhibits the correct qualitative behavior,
the detailed slopes are incorrect.  An explanation for this
is that clusters likely will not simply become rounder, in the case of
a single component cluster, or have unchanging clumps approaching
each other in the case of a bimodal cluster.  Up to a point, general 
relaxation tends to concentrate mass distributions as well so that 
the parameters describing the mass distributions are likely to evolve
also.  As an illustration, consider the path followed in power 
ratio space if a single $\beta$ model cluster with $\beta=0.75$, 
$a_{core}=0.3~\rm Mpc$, $\epsilon_x=0.6$ evolves to a cluster 
with smaller $\epsilon_x=0.3$, but also smaller $a_{core}=0.1~\rm Mpc$.
This is given by joining the topmost open circle with the topmost 
filled circle of Figure \ref{powrat42.fit}.  The slope is a better 
match to that of the evolutionary track suggesting that indeed
cluster evolution is accompanied by specific changes in cluster 
parameters, such as the core radius and the ellipticity. This 
issue is considered in more detail in \S\ref{clusevol.simple}.

\subsection{Statistical Comparisons at Low Redshift ($z <0.1$)
\label {SClow1}}

The distribution of clusters along the evolutionary track is
determined by the balance between the cluster relaxation and formation
rates, and hence isolates the information which likely specifies
$\Omega$. In keeping with our phenomenological approach (see BTb), we
make the following intuitive definitions of these rates guided in part
by results from the preceding section.  We consider the ``relaxation
rate'' to be defined as the hypothetical rate at which clusters would
move down the evolutionary track in isolation; i.e. when undisturbed
by mass accreting from outside the specified aperture.  The
``formation rate'' consists of two components, a ``birth rate'' and
``merger rate''. The birth rate is the rate at which clusters first
appear on the upper parts of the track immediately after forming out
of the background matter. In contrast, the ``merger rate'' is the rate
at which a mature cluster jumps from a lower position to a higher
position on the track due to the accretion of mass from outside the
specified aperture.  If these rates as described in terms of movement
along the evolutionary track are in fact physically similar to typical
theoretical constructions (e.g., Richstone et al. 1992; Lacey \& Cole
1993; Bartelmann, Ehlers, \& Schneider 1993; Kauffmann \& White 1993;
Nakamura, Hattori, \& Mineshige 1995), then this would suggest that to
probe $\Omega$ effectively we should consider the location of clusters
in power-ratio space relative to a set of rotated coordinates where
one axis lies along the correlation line.

In order to make use of all the information provided by the power
ratios, it is also important to consider the distribution along an
axis perpendicular to the evolutionary track. In essence, for a given
point on the evolutionary track a cluster lying perpendicularly to the
left of the track possesses a larger amount of smaller scale
structures (i.e. excess $P_3/P_0$ or $P_4/P_0$) but a less pronounced
aggregate, larger scale structure (i.e. less $P_2/P_0$); the opposite
is true for clusters lying perpendicularly to the right of the track.
Hence, the distribution of clusters perpendicular to the evolutionary
track is probably not intimately related to the rates of relaxation
and formation described above and is thus unlikely to be strongly
dependent on $\Omega$. However, the perpendicular distribution may
provide different, yet potentially equally interesting, constraints on
the structure of clusters.  For example, the perpendicular
distribution would seem to be related to the shape of the power
spectrum of initial fluctuations on cluster scales (if that
information is preserved during cluster evolution), or to the manner
of cluster relaxation (such as by gravitational/pressure
forces). Whether such effects are indeed important will require
further study (see Buote \& Xu 1996).  Finally, the slope and
intercept of the correlation line itself may be related either to the
specifics of cluster evolution or to the cosmological model.

Figure \ref{powrat42.fit} shows the ``best measured'' clusters from
the ROSAT sample of BTb (error bars are excluded to avoid clutter).
These clusters are defined to have error bars that span less then a
decade except for those with $P_4/P_0$ or $P_2/P_0$ values $\le
0.25\times 10^{-7}$.  This gives a sample of 31 clusters.  The light
solid line shows the best fit line to these data and is given by ${\rm
log}(P_4/P_0)= a + b~{\rm log}(P_2/P_0)$ with $a=-0.86\pm 0.4$ and
$b=1.188\pm 0.077$ where standard errors are indicated. This best-fit
line is adopted for the evolutionary track. (Note that the parameters
of this best-fit correlation line and those that follow are changed
negligibly when instead all clusters corresponding to the augmented
Edge et al. 1990 sample are used in the fits -- Buote \& Xu 1996.)

We transform the coordinates from the power ratios
$(P_2/P_0,P_4/P_0)$ (specifically, the log of the power ratios)
to a system $(x_p, y_p)$ oriented such that the $x_p$ axis coincides
with the best-fit line above and the $y_p$ axis is perpendicular to
$x_p$.  The origin of the new coordinates is set to ${\rm log} 
(P_2/P_0)=-8.0$ and ${\rm log} (P_4/P_0)=-10.37$.
Figure \ref{histo420} shows the distribution of clusters when referred
to the new coordinate system where $x_p$ and $y_p$ are both measured
in log units.  The distributions for two sets of data samples are
shown.  Out of a total of 59 clusters considered, 44 had power ratios
that could be determined for a 1 Mpc aperture (see BTb).  This sample
is shown by the heavy solid histogram.  The distribution of the ``best
observed'' 31 of these clusters is given by the light solid histogram
and that of the simulated clusters by the dashed histogram.

We construct the simulated cluster sample by combining the power
ratios of the three projections (see \S \ref{method}) of each of the
simulated clusters for the two lowest redshifts considered,
$z=0$ and 0.07, which correspond approximately to those of the
observed cluster sample.  This assumes that each projection of a given
cluster represents an independent observation for the purposes of
statistical analysis, i.e., each simulated cluster represents the
``average'' cluster of that size and the process of observing is
simply viewing that cluster from random angles.  This is standard
practice in those cases where small numbers of simulated clusters are
available (e.g., Evrard et al. 1993, Mohr et al. 1995).  We further
assume that a given cluster can be considered independent at two
different (but close) times in its evolution.  Although each simulated
cluster originates from a single fluctuation in the initial matter
distribution, every cluster passes through a similar series of
morphological stages during its evolution (based on the simulations).
Hence the observation of a number of clusters of the same mass is
equivalent to observing one cluster but picking it out at several
different stages in its evolution.  Although there are still really
six independently simulated clusters, provided the forgoing
assumptions are not
grossly inaccurate, the augmented
sample of simulated clusters will allow for interesting quantitative
comparison with the sample of ROSAT clusters. In any case, the
conclusions drawn below do not change if only one specific redshift is
considered.

In constructing the sample of simulated clusters, equal weight is
assigned to each cluster.  Ideally, we must ensure that the contents of
the model sample agree with constraints from the luminosity function
and the limiting flux of the ROSAT sample we use for comparison.  We
find, however, that in a flux-limited sample corresponding to the
ROSAT data and the observed luminosity function of ROSAT clusters
(e.g., Briel \& Henry 1993) the abundances of clusters with masses
represented by the simulated clusters will indeed be uniform.  Hence
the assignment of equal weights to each of the clusters does indeed
give a distribution of clusters (in mass) that approximately
replicates the observed distribution.

The distributions for the data and for the simulations in the $x_p$
direction are inconsistent. The simulated clusters are distributed
more uniformly over the evolutionary track than the ROSAT sample and
exhibit an excess of large $x_p$ clusters; i.e. the simulated clusters
have more substructure on the indicated scale ($\sim 1$ Mpc) than is
found in the data. As a quantitative measure of this
disagreement, a Kolmogorov-Smirnov (KS) test applied to the observed
and simulated samples (again with the simulated sample constructed 
as described above) gives a probability of P(KS)=4.4\% that the two
samples arise from the same distribution.  

This result appears similar to the result of one of the statistics
used by Mohr et al. (1995) who found that their measures of axial
ratios were more broadly distributed for a sample of Einstein IPC
clusters than for the $\Omega = 1$ simulations. However,
interpretation of the Mohr et al. result is unclear because the same
cluster scales are not consistently compared -- see BTa \S 6. for a
discussion. (The other statistics used by Mohr et al. (1995), the
centroid-shift and radial fall-off, were found to be essentially {\it
consistent} for the $\Omega=1$ simulated clusters and IPC X-ray
clusters.)

Assessing the sensitivity of this result to the incompleteness of the
simulated cluster sample cannot be definitely answered with present
simulations.  However, a study in progress with purely dissipationless
simulations shows that qualitatively the above conclusions are still
obtained when a statistically large and independent sample of clusters
is used (Buote \& Xu 1996).  Here, we consider the sensitivity of this
conclusion to incompleteness in the data sample.  The observed
clusters comprise a sample which is $\sim 50\% - 60\%$ complete for
the brightest 35 clusters (BTb).  Clusters with fluxes above the
limiting value were missed for various reasons including instrument
mispointings, images that were too large to fit inside the PSPC ribs
(see BTb), and because they simply were not observed with the PSPC.
Since clusters were not excluded for reasons related to their
morphology, the missing clusters should have a distribution of power
ratios similar to the observed sample.  This expectation is supported
by available information on the missing clusters.  For example, the 1
Mpc aperture for some of the clusters in the data sample did not fit
within the inner ring of the PSPC. In many cases, however, a
significant fraction of the 1 Mpc aperture ($\gtrsim 0.75$ Mpc) did
fit within the ring which allows us to approximate the 1 Mpc
value. Also, a qualitative assessment of the cluster structure is
possible by visual examination of the emission in the 1 Mpc aperture
lying outside the inner ring (BTb).  Some of the clusters not observed
by the PSPC were observed by the {\it Einstein} Imaging Proportional
Counter (IPC) from which the cluster morphology can also be
examined. Using both power ratios in apertures of radii $0.75 \lesssim
R_{ap} < 0.9$ Mpc and/or visual examination of available data for the
missing clusters we find that not only is the distribution of power
ratios similar to that shown in Figure \ref{histo420}, but that only
$\sim 2$ out of $\sim 40$ clusters have power ratios in the region
$x_p\ga 4.5$ where the disparity with simulated clusters is most
severe.

A further demonstration of the stability of the distribution of
observed clusters is provided by the sample of ``best measured
clusters.''  These were not selected on the basis of their morphology
but only on the size of the random errors associated with the
observation.  The distribution of these clusters should be similar to
that of the larger sample, which is demonstrated by Figure
\ref{histo420}. Similarly, if we consider only the $\sim 35$ brightest
clusters which comprises a sample that is $\sim 50\% - 60\%$ complete
(see BTb) then we obtain qualitatively similar distributions.  Again
as a simple measure of this similarity, we get P(KS)=3\% in the $x_p$
direction, consistent with the results for the whole sample.

The distributions in the $y_p$ direction are consistent as indicated
by the large probability P(KS)=32\%; again, this result holds for
subsets of the data sample where, e.g., the $\sim 35$ brightest
clusters give P(KS)=55\%.  Intuitively this result implies that for a
particular dynamical age (i.e. at a particular position along the
evolutionary track) simulated clusters in the $\Omega=1$ simulation
tend to ``look'' like real clusters. This correspondence may represent
a success of the cosmological model (see above and \S \ref{disscon}).

The $P_3/P_0$ -- $P_2/P_0$ plane can be examined in the same way as
above.  We fitted a straight line to the evolutionary track given by
the ``best measured'' clusters of Figure \ref{powrat32} resulting in
${\rm log}(P_3/P_0)= a + b~{\rm log} (P_2/P_0)$ where $a=-1.81\pm
0.58$ and $b=0.991\pm 0.11$ with standard errors.  As before, define
the $x_p$ and $y_p$ axes to lie along and perpendicular to the
best-fit line, respectively, and set the origin of the new coordinate
system at ${\rm log}(P_2/P_0)=-8.0$ and ${\rm log}(P_3/P_0)=-9.07$.
The distribution of clusters relative to $(x_p, y_p)$ are shown in
Figure \ref{histo320} where the coordinates are measured in log units.

The results in this case are consistent with those from the $P_4/P_0$
-- $P_2/P_0$ correlation. First, there is again an excess of simulated
clusters for large $x_p$ values; i.e. there is generally too much
substructure in the simulated clusters. However, the inconsistency of
the distributions in the $x_p$ direction is much more significant than
in the $P_4/P_0 - P_2/P_0$ case; i.e. KS probability of only 0.15\%
($>3\sigma$) that the simulated and ROSAT clusters are
consistent. (Note that this inconsistency involving an ``odd moment''
differs markedly from the centroid-shift results from Mohr et
al. 1995.)  In the $y_p$ direction the distributions are consistent at
the same level as before with KS probability P(KS)=37\%; i.e. again,
individual clusters of similar dynamical ages (position along the
evolutionary track) ``look'' like real clusters in terms of their
morphology.

The conclusions from the the $P_3/P_0$ -- $P_2/P_0$ correlations are
not independent of those from the $P_4/P_0$ -- $P_2/P_0$ correlation
because they both involve $P_2/P_0$.  However, the two sets of
correlations give complementary information.  The power ratio
$P_3/P_0$ is particularly sensitive to bimodal cluster structure (with
clumps separated by $r_{sep}\la$ the aperture size) where the
secondary component is smaller than the primary component ($P_3/P_0=0$
for exactly equal-sized spherical components since the current set of
ratios are centered on the centroids of the images -- BTa).  On the
other hand, the even moments are particularly sensitive to the overall
ellipticity or ``boxiness'' of clusters.  The ratio $P_2/P_0$ will be
large for highly elongated structures such as flattened single
component clusters and bimodals with comparably sized components.  The
very significant deficit of high $x_p$ clusters in the $P_3/P_0$ --
$P_2/P_0$ case indicates that there are far too many bimodal clusters
with smaller secondaries in the simulations as compared to the data.
The somewhat less significant deficit of large $x_p$ clusters in the
$P_4/P_0$ -- $P_2/P_0$ case indicates a less numerous surplus in
highly elongated single component clusters and equally sized bimodals.

We now consider $P_1^{(pk)}/P_0^{(pk)}$ which is defined so that the
aperture is centered at the peak of $\Sigma_x$. By construction
$P_1^{(pk)}/P_0^{(pk)}$ is not independent of the $P_m/P_0$, however,
it is useful to consider power ratios centered on the peak emission
because they are particularly sensitive to bimodal clusters with
comparably sized components while being insensitive to ellipsoidal
single component clusters.  Figure \ref{histop10} shows the
distribution of clusters in $P_1^{(pk)}/P_0^{(pk)}$.  The surplus of
bimodal clusters suggested by the $P_4/P_0$ -- $P_2/P_0$ results is
confirmed again by the excess in large $P_1^{(pk)}/P_0^{(pk)}$
simulated clusters.  The KS probability for consistency is P(KS)=1.2\%
when all observed clusters are considered. This result disagrees with
the centroid-shift results of Mohr et. al. (1995).

\subsection{Statistical Considerations for Higher Redshift ($z >0.1$)
\label {SChigh1}}

At present, there is a paucity of X--ray data available for
high-redshift ($z\ga 0.2$) clusters (e.g., Donahue \& Stocke 1995,
Castander et al. 1994).  Unfortunately, this situation will persist
until the launch of AXAF which will finally enable higher redshifts to
be significantly probed.  However, comparison of high redshift
simulated clusters with both the observed ROSAT sample and low
redshift simulated clusters allows the study of cluster evolution in
the $\Omega=1$ CDM cosmogony.  A principle consideration is how the
cluster formation rate (i.e. $x_p$ distribution -- see beginning of \S
\ref{SClow1}) varies with redshift and whether the results show the
strong evolution expected from analytic studies (e.g., Richstone
et. al. 1992).  We also suggest potential observational tests of the
CDM cosmogony with high $z$ clusters.

The $z>0.1$ simulated clusters can be treated in the same manner as
the low redshift clusters.  Each cluster is viewed down three random,
orthogonal axes and the power ratios are plotted in the manner of
Figures \ref{powrat42} and \ref{powrat32}.  This method of generating
the sample of high redshift simulated clusters is not entirely
consistent since we are considering the same clusters at high and low
redshift.  We might expect that since individual clusters evolve in
different ways, the sample that should be compared with observations
will not consist of the same clusters at different redshifts.  As with
the incompleteness of the simulated sample discussed in the previous
section, we cannot usefully address this issue at present and thus our
following analysis is intended only to be suggestive.

The location of clusters can be expressed in terms of the rotated
coordinates $(x_p, y_p)$ where we assume the same set of coordinates
used in the low $z$ cases in order to have consistent comparisons.
The distribution of clusters in the $P_4/P_0$ -- $P_2/P_0$ plane are
shown in Figure \ref{histo42z} and the distribution in $P_3/P_0$ --
$P_2/P_0$ are shown in Figure \ref{histo32z}.  Three redshift ranges
are considered.  The range $z_1$ consists of simulated clusters at
$z=0.14$ and 0.23, the range $z_2$ is composed of clusters at $z=0.35$
and 0.49, and the range $z_3$ is given by the clusters at
$z=0.68$. Higher redshifts are not considered because some of the six
NFW clusters begin to disperse by $z>0.68$.

Consider first the $P_4/P_0$ -- $P_2/P_0$ plane.  Similar to what we
found for the low-redshift simulated clusters (see \S \ref{SClow1}), 
the high-redshift simulated clusters are roughly evenly distributed in 
the $x_p$ direction with a sizeable number of large $x_p$ clusters.
KS tests of the consistency of low and high redshift simulations
yield P(KS:$z_1$, $z_{low}$) = 83\%, 
P(KS:$z_2$, $z_{low}$) = 35\%, and P(KS:$z_3$, $z_{low}$) = 15\%.
This indicates that the intermediate $z$
simulated clusters $z_1$ and $z_2$ have $x_p$ distributions consistent
with that of $z_{low}$, while $z_3$ is in marginal
disagreement. These relationships are further supported by comparing
the high-redshift simulated clusters with the observed ROSAT clusters 
which are all at low $z$: P(KS:$z_1$, data)=4.9\%, P(KS:$z_2$, data)
=12\%, and P(KS:$z_3$, data)=0.55\%.  (Again, the highest redshift
clusters seem to have somewhat different properties from the low
redshift simulated clusters.)  The
distribution of high redshift simulated clusters in the $y_p$
direction is consistent with the data, except again for the highest
redshift case.

These results demonstrate that the morphologies of the sample of
simulated clusters, as measured by the even power ratios, remains
essentially unchanged over the redshift range from the present to
$z\sim 0.6$. The equivalence of the $x_p$ distributions means
that by $z\sim 0.6$ the rate of clusters arriving in the upper regions
of the evolutionary track has come into
balance with the relaxation rate. If we assume that the relaxation
rate does not depend strongly on the cosmological model or on the
epoch being considered (i.e., the self gravity of cluster material
dominates both tidal forces from surrounding large-scale structures
and the effects of an expanding universe), the formation rate of
clusters must then be constant.

Whether this balance between formation and relaxation after a specific
redshift is a common feature of cluster formation scenarios or instead
depends on the specific cosmogony requires consideration of a variety
of models.  However, this result is not clearly anticipated by
semi--analytic considerations of structure formation such as those by
Richstone et al. (1992).  The constancy in the distribution of cluster
morphologies and the apparent change at $z\sim 0.6$ provides a
new test of the model.  Current X--ray observations 
are beginning to probe this
redshift range, but the usefulness of the data is limited by poor
photon statistics and by the difficulty of identifying clusters at
these redshifts.  Hopefully future high resolution observations by
AXAF will sample this regime with some completeness.

If correct, the change in the distribution of power-ratios at
$z\ga 0.6$ is of interest.  At low redshifts, clusters are
distributed evenly in $y_p$ about the best fit line to the data.  The
$z_3$ clusters, however, are strongly skewed so that many more have
positive values of $y_p$.  These clusters sit above the correlation
line in Figure \ref{powrat42.fit} and would be considered relatively
unrelaxed with significant small-scale structure.  This is supported
by the the distribution along the $x_p$ direction which shows an
excess of clusters with very large values of $x_p$.  These properties
identify $z\gtrsim 0.6$ as an epoch where clusters are rapidly
forming out of background matter and where the relaxation rate is
insufficient to balance the formation rate.  The formation rate levels
off in the epochs immediately after this time.

The cluster distributions in the $P_3/P_0$ -- $P_2/P_0$ are consistent
with the above behavior although with somewhat more uncertainty.  The
$z_3$ clusters are again skewed toward high $x_p$ values relative to
those at lower redshifts and the distribution in $y_p$ indicates an
overpopulation at positive $y_p$.  The time around $z\sim 0.6$
is also identified as the era of rapid cluster formation.  Clusters
at $z_1$, however, also show a shift to large $y_p$, although it is
less significant than at high redshift and the shift is into the bin
just adjacent to $y_p=0$.  This may be the result of a statistical
fluctuation since there are $\sim 36$ measurements of the power ratios
at $z_1$.  We could have just picked a particular time which happens
to give a $\la 2\sigma$ fluctuation to larger $y_p$.  Alternatively,
the number of clusters with large $P_3/P_0$ could really be high at
certain redshifts.  This issue cannot be settled with the present
limited simulations.

The distribution of $P_1^{(pk)}/P_0^{(pk)}$ for the higher redshifts
is plotted in Figure \ref{histop1z}.  For the $z_1$ and $z_2$
clusters, the results are consistent with those from the other power
ratios.  However, the distribution shifts towards {\it smaller}
$P_1^{(pk)}/P_0^{(pk)}$ at the highest redshifts considered.  At
first, this may appear to contradict the picture of cluster evolution
outlined above, but it actually provides an interesting additional
constraint while highlighting the utility of the dipole power ratio.
Recall that $P_1^{(pk)}/P_0^{(pk)}$ is large for bimodal clusters with
roughly equally sized components but small for elongated single
component clusters.  In contrast, the centroided even power ratios are
large for both sets of clusters (BTa).  The small number of large
$P_1^{(pk)}/P_0^{(pk)}$ clusters at high $z$ and the excess of
positive $y_p$ clusters in the $P_4/P_0$ -- $P_2/P_0$ plane implies
that there are many rather unrelaxed single component clusters, but
very few large bimodal clusters; here ``single component'' means an
isolated but possibly morphologically complex lump of material.
This is a reasonable situation.  At the earliest times, single lumps
are forming out of the background matter distribution.  These are
unrelaxed so have large centroided power ratios; the distribution
is skewed towards high $x_p$ and positive $y_p$.
However, the bimodals
can only occur after two of these nascent lumps have collided and
begun to merge.  This occurs some time after the initial formation of
the single lumps.  Thus, the epoch around $z\sim 0.6$ is a the
period when virializing single component clusters are rapidly emerging
from the background matter.  After this time, clusters form primarily
by mergers with other lumps (there are few single lumps emerging from
the background since the distribution in $y_p$ is evenly spread
between positive and negative values) and it is this rate which
balances the rate of relaxation and results in the constant
distribution of clusters along the $x_p$ direction of the power ratio
correlations.  It is evident that the cluster formation rate is not
simply the rate of making bound mass concentrations out of initial
density fluctuations, but also involves knowing the merger rate at the
given time.

\subsection{Constraints on Cluster Parameters from the Evolutionary 
Track}\label{clusevol.simple}

\begin{figure}
\caption{  \label{clus1z} }

\plotone{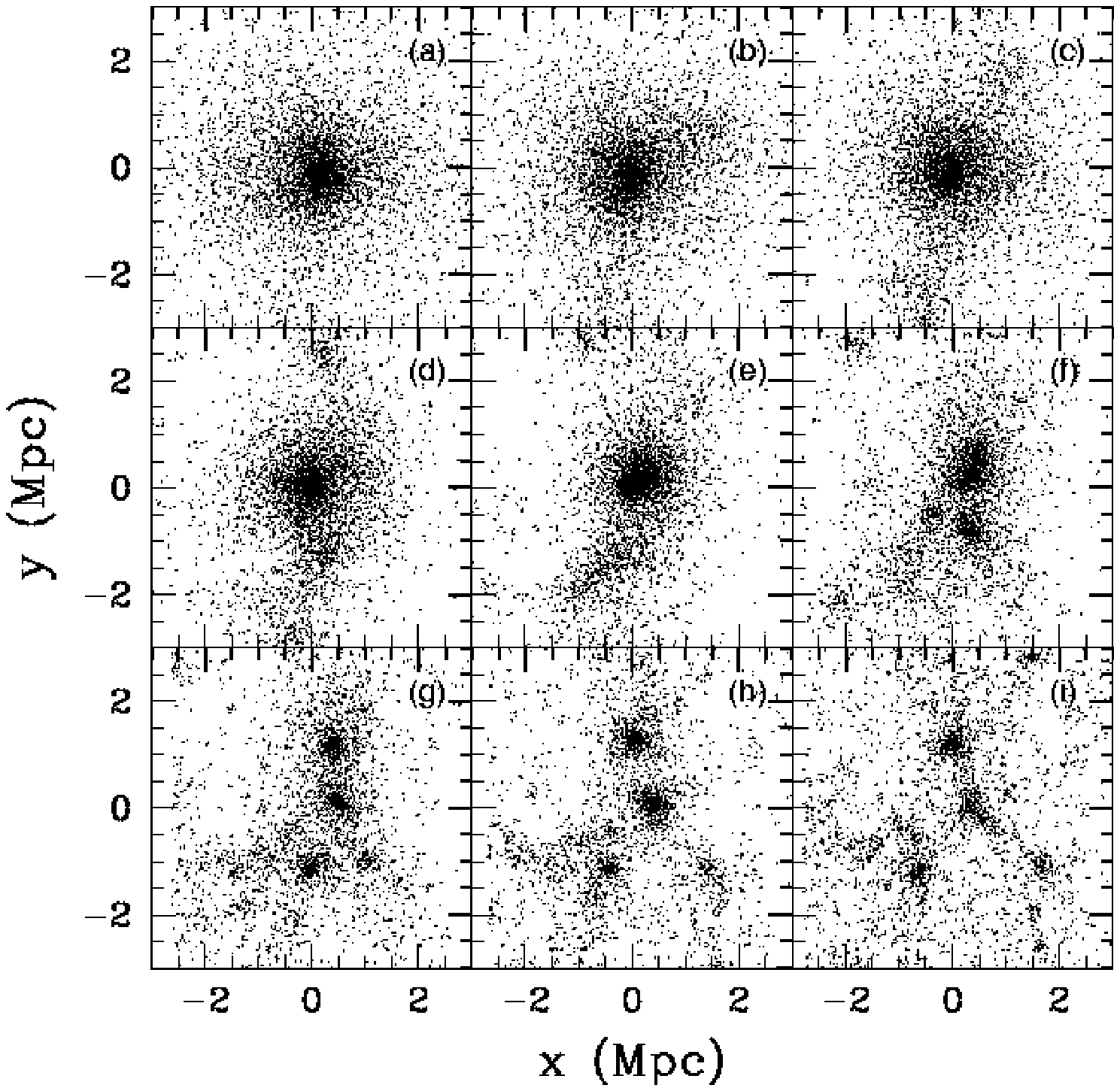}

\raggedright
The SPH gas particles for cluster CL1 are shown viewed down the $z$ axis
for the same set of redshifts as Figure \ref{clus1x}.
\end{figure}

\begin{figure}
\caption{  \label{powrat42} }

\plottwo{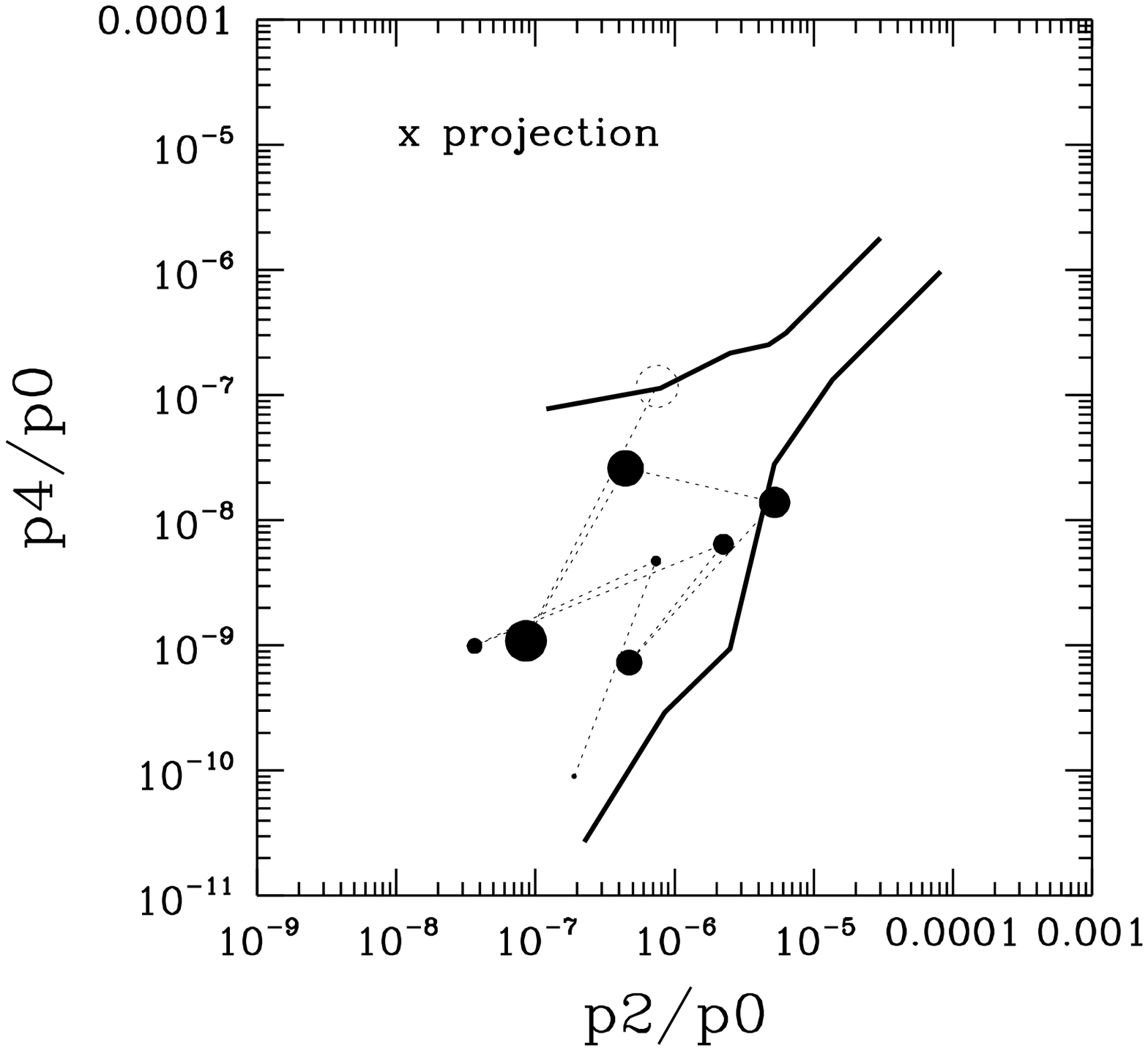}{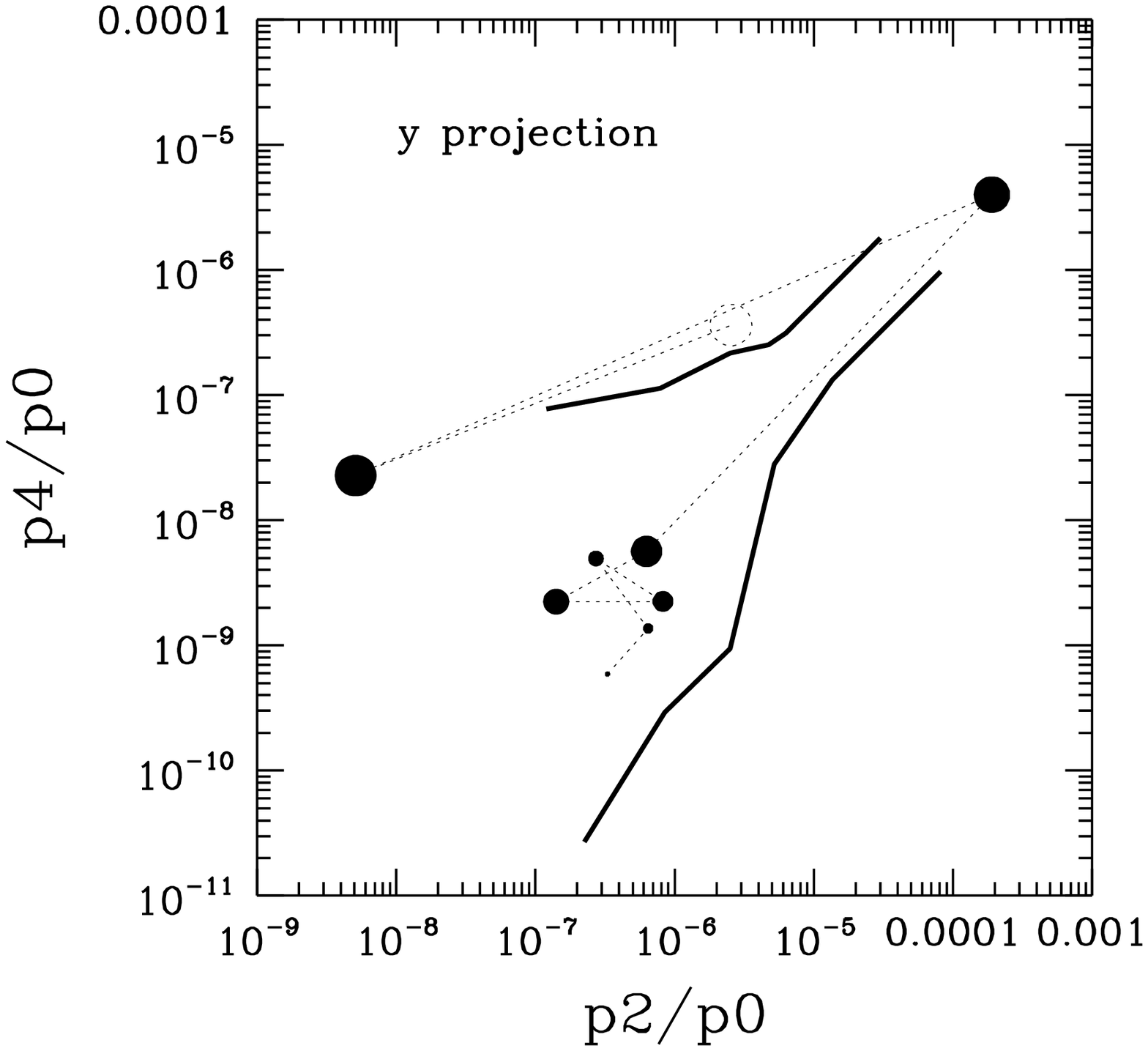}

\plottwo{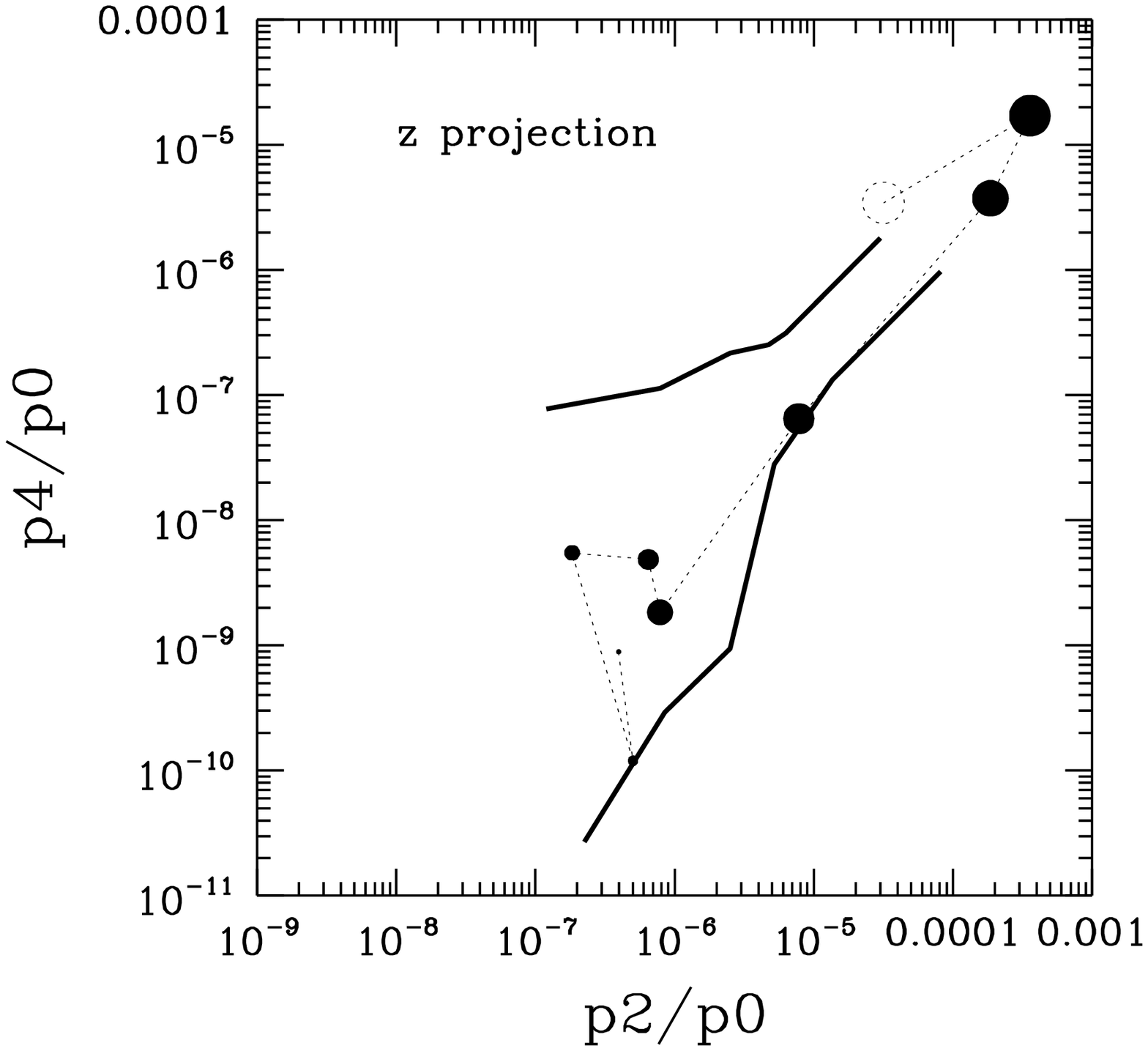}{blank.ps}

\raggedright
The locations of the largest cluster (CL1) in the plane formed by $P_4/P_0$
and $P_2/P_0$ are given for, (a) the $x$ projection, (b) the $y$ projection,
and (c) the $z$ projection.  An aperture of 1 Mpc is assumed.  The dashed 
empty circle corresponds to the
earliest redshift, $z=1.4$.  The filled circles show smaller redshifts
where the size of the circle decreases with redshift.  In decreasing
size, the circles correspond to $z$=0.95, 0.68, 0.49, 0.35, 0.23,
0.14, 0.065, and 0.  The circles are connected by a dashed line to emphasize
the motion of the cluster in power ratio space.  The heavy solid lines
indicate approximately the region occupied by the observed power ratios 
from BTb.
\end{figure}

\begin{figure}
\caption{  \label{powrat32} }

\plottwo{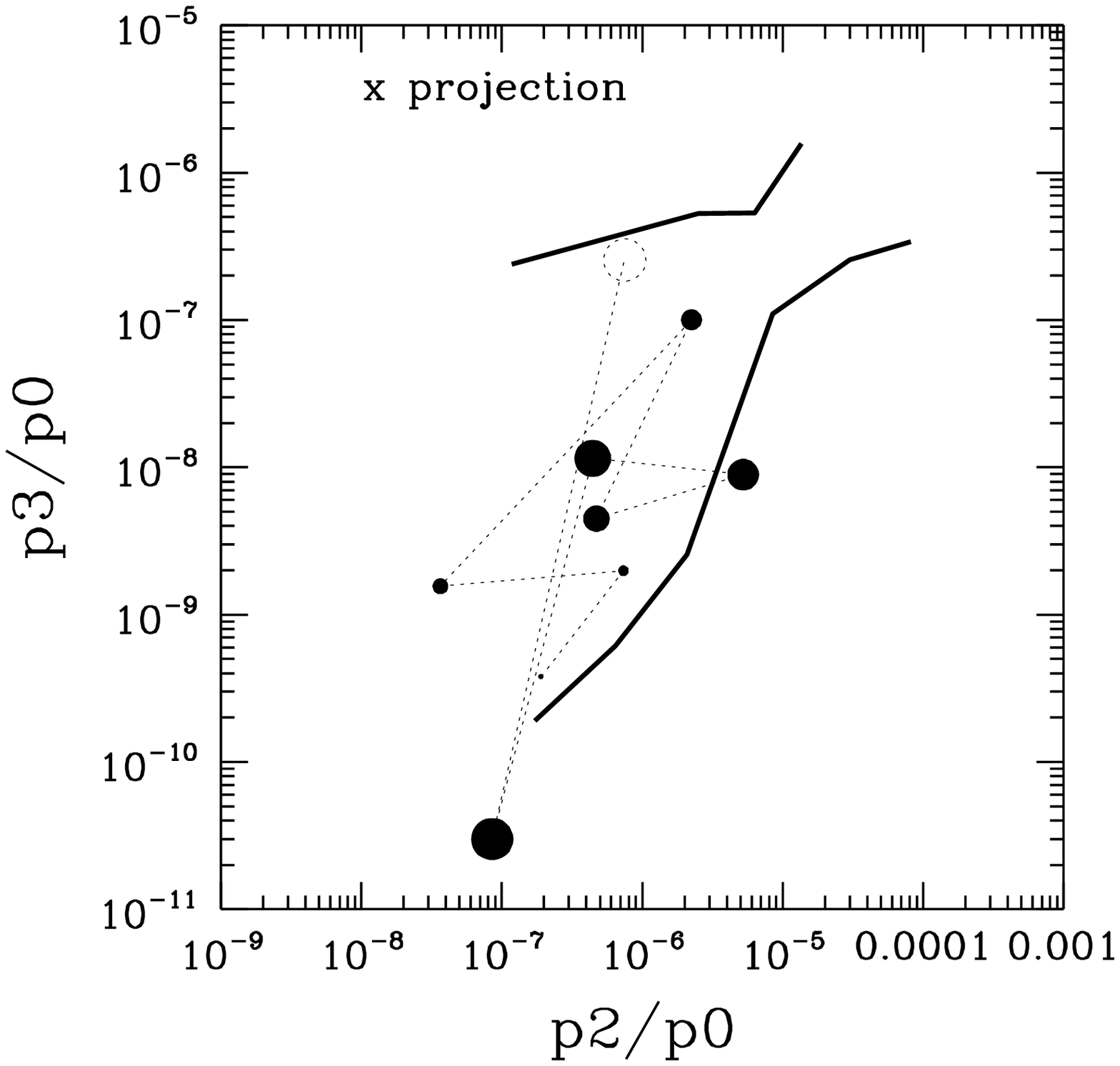}{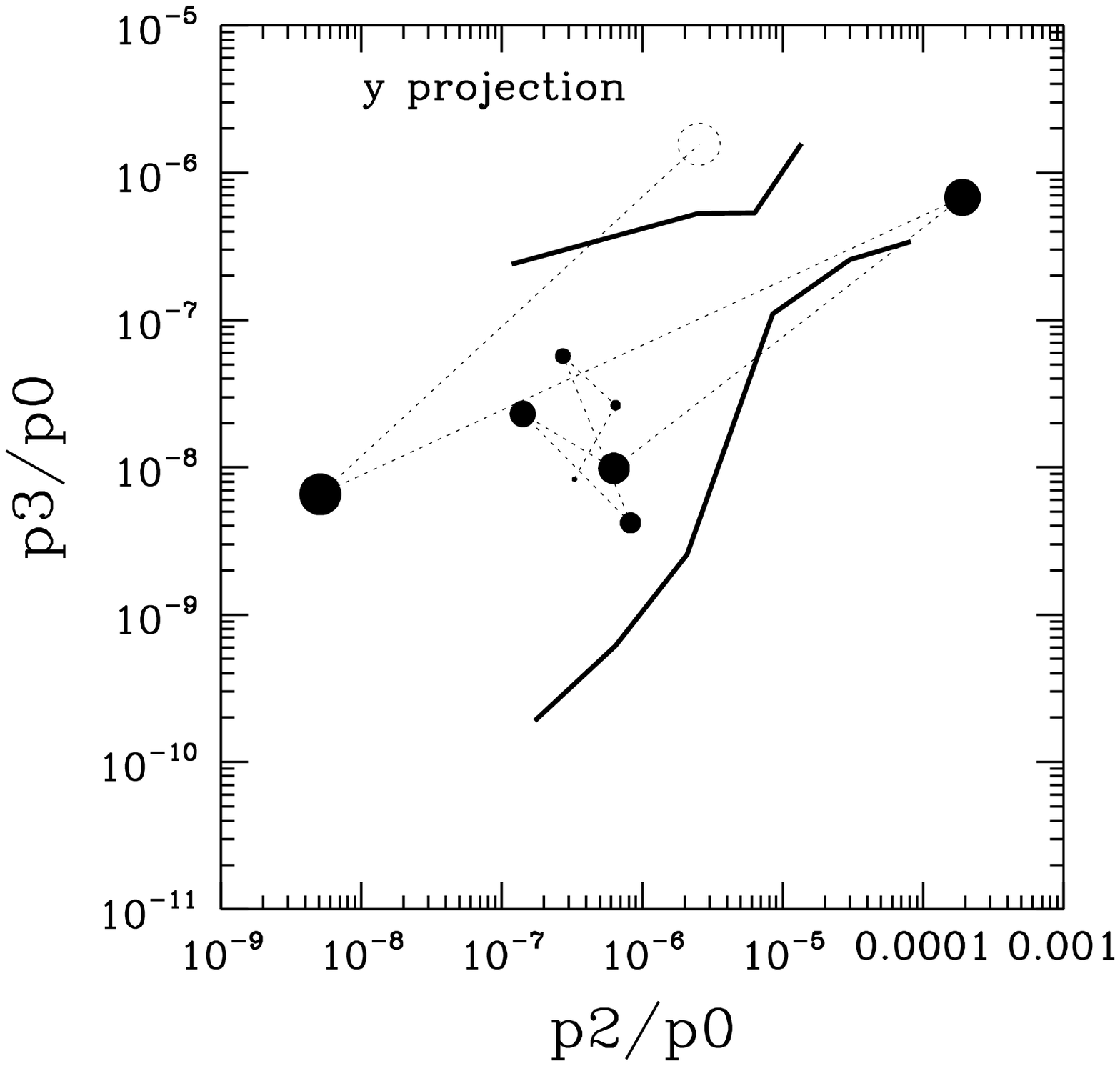}

\plottwo{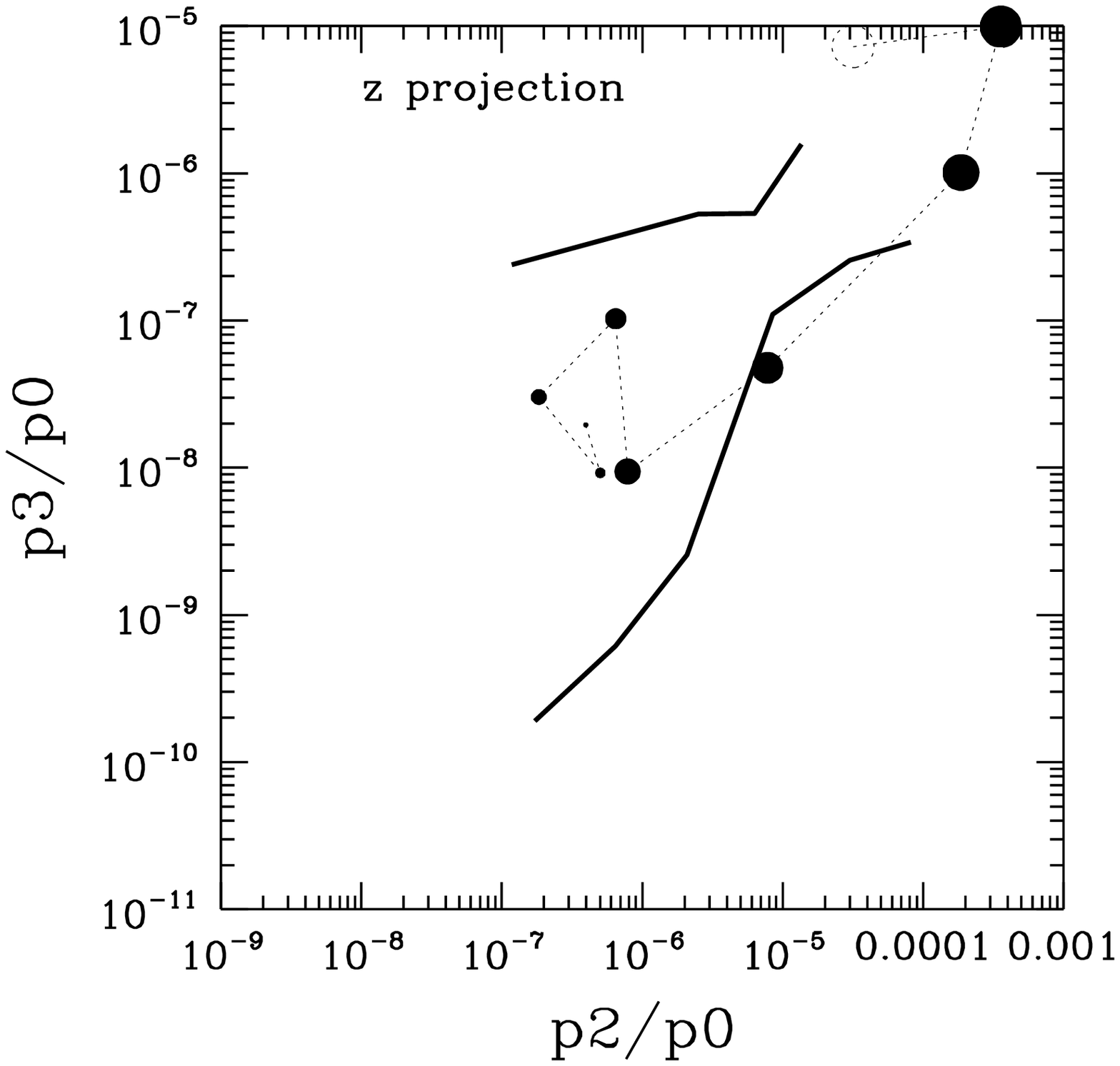}{blank.ps}

\raggedright
The locations of the largest cluster (CL1) in the plane formed by $P_3/P_0$
and $P_2/P_0$ are given for, (a) the $x$ projection, (b) the $y$ projection,
and (c) the $z$ projection.  The points are plotted as in Figure 
\ref{powrat42}.  The heavy solid lines again shows the approximate
region occupied by observed values from BTb.
\end{figure}

\begin{figure}
\caption{  \label{powrat42.fit} }

\plotone{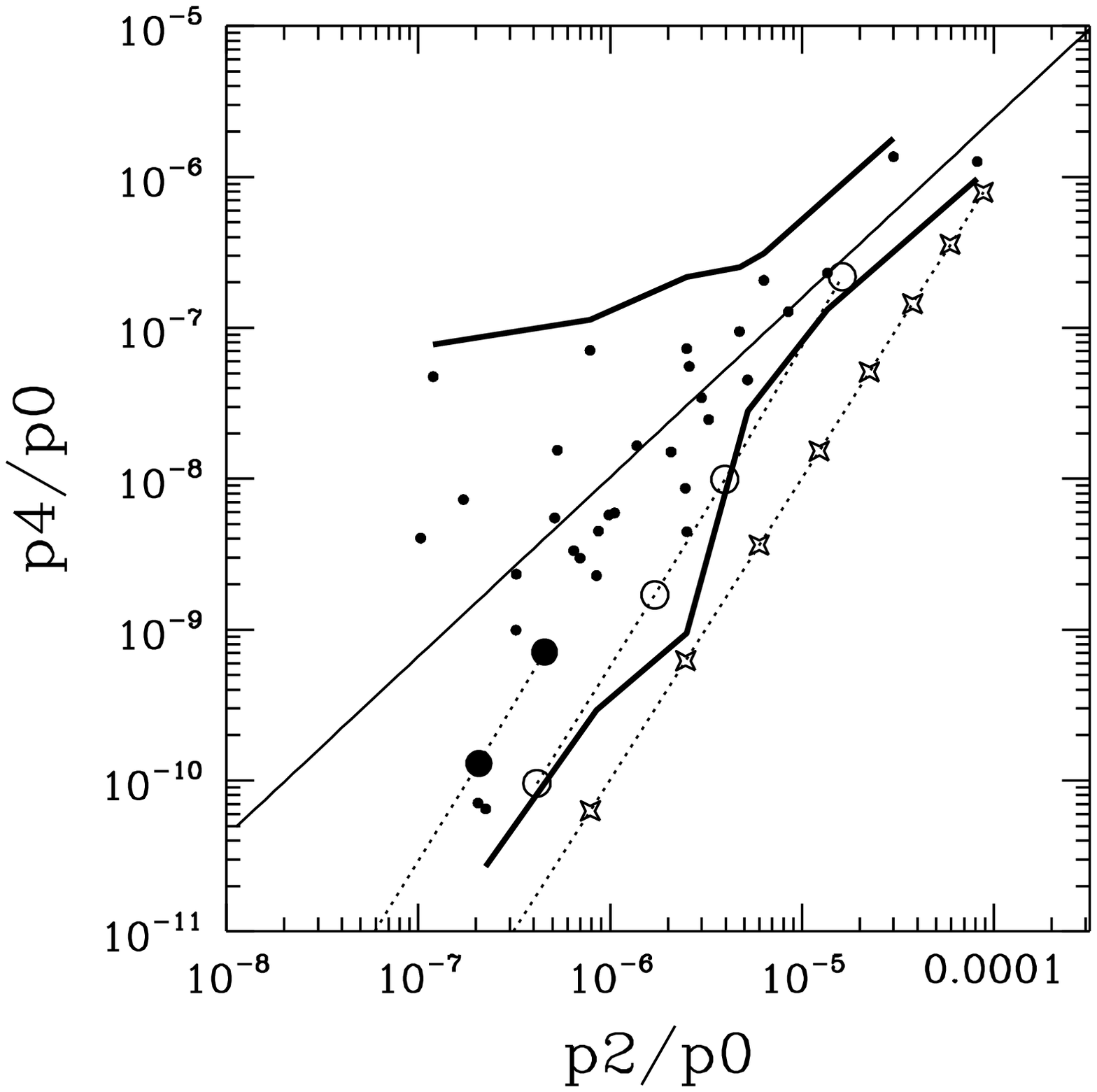}

\raggedright
The power ratios $P_4/P_0$ are plotted versus $P_2/P_0$.  The small
filled circles correspond to the ``best measured'' clusters in the 
sample of BTb.  The heavy solid lines correspond roughly to the range of
observed values and the light solid line is the best fit line to
the cluster data.  The large open circles are the power ratios computed
for single component $\beta$ models with fixed core radius $a_{core} = 
0.3~\rm Mpc$ and $\beta=0.75$.  The topmost open circle corresponds
to an ellipticity of $\epsilon_x = 0.6$.  Proceeding downwards,
the circles correspond to $\epsilon_x = 0.3, 0.2,\rm and~ 0.1$,
respectively.  The large filled circles correspond to single $\beta$ 
models with $\beta=0.75$, $a_{core} = 0.1 ~\rm Mpc$.  The topmost circle 
has $\epsilon_x = 0.3$, and the bottom one has $\epsilon_x = 0.2$.
The stars give the power ratios for bimodal clusters where each
component is a single $\beta$ model with $\beta=0.75$ and $a_{core} =
0.3~\rm Mpc$.  The assumed separation of the components is $a_{sep}=
1~\rm Mpc$ for the topmost star and $r_{sep}= 0.9, 0.8, 0.7$, etc. for
successively lower stars.  Circles and stars are joined with dotted 
lines to indicate the implied linear relation of $P_4/P_0$ to $P_2/P_0$.
\end{figure}

\begin{figure}
\caption{  \label{histo420} }

\plottwo{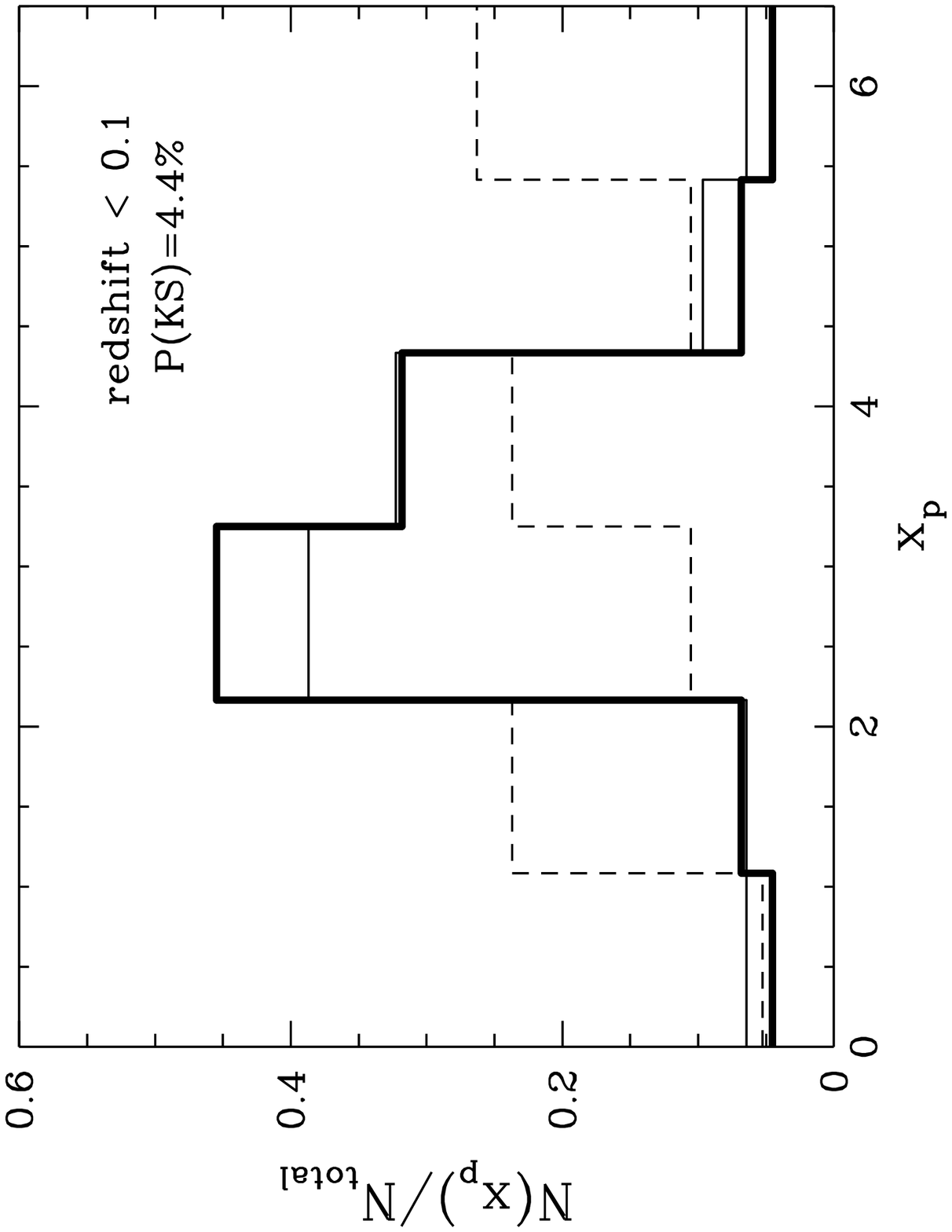}{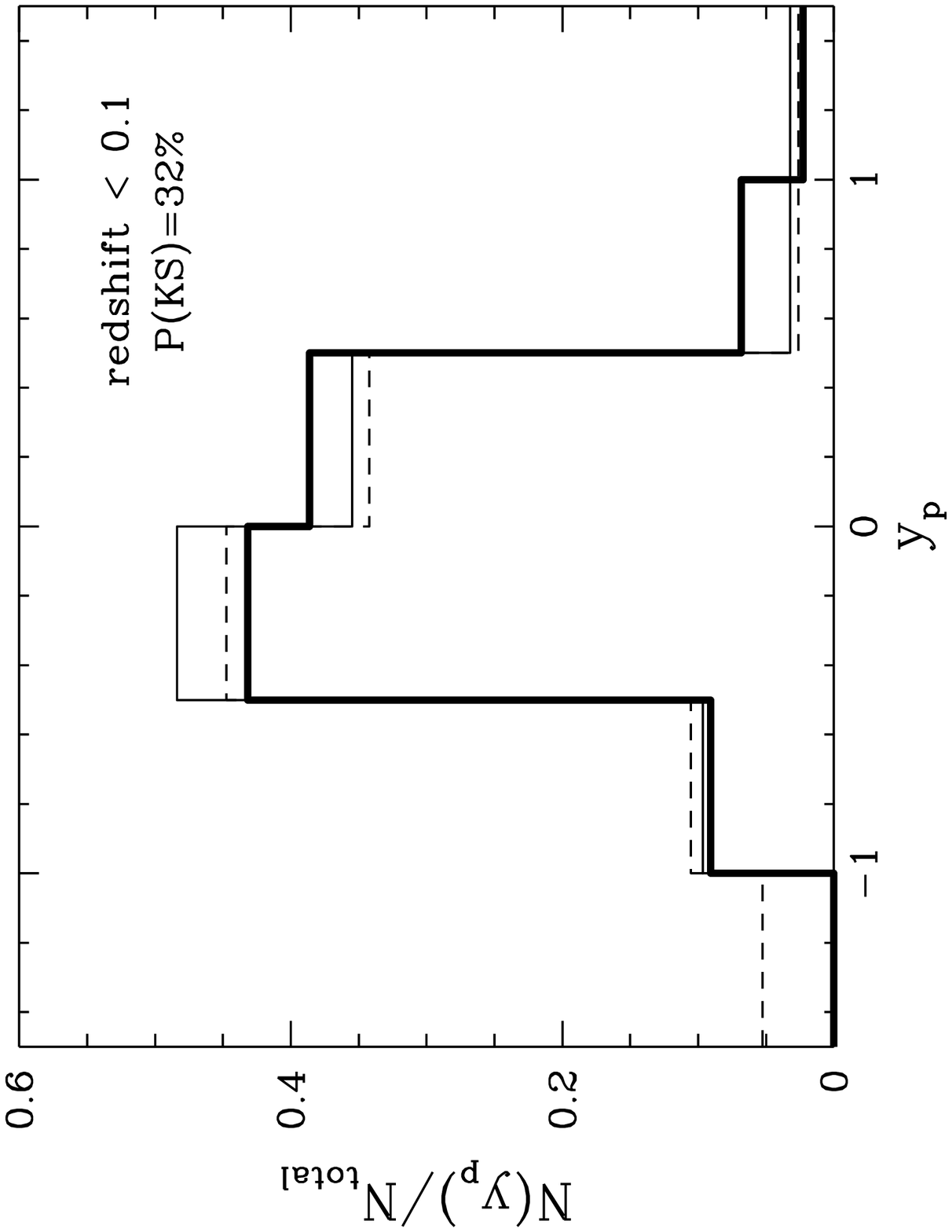}

\raggedright
The distribution of clusters in the $P_4/P_0$ -- $P_2/P_0$ plane are
shown.  Panel (a) shows the number (normalized to the total number
of clusters in the given sample) of clusters with given
$x_p$ (defined in the text).  Panel (b) shows the number (normalized
to the total) with given values of $y_p$.  The heavy solid histogram
corresponds to the complete observed sample of 44 clusters and the light
histogram refer only to the ``best measured'' cases, numbering 31
clusters.  The dashed histogram is the distribution of simulated clusters
for the two lowest redshifts of $z=0$ and $0.07$.
\end{figure}

\begin{figure}
\caption{  \label{histo320} }

\plottwo{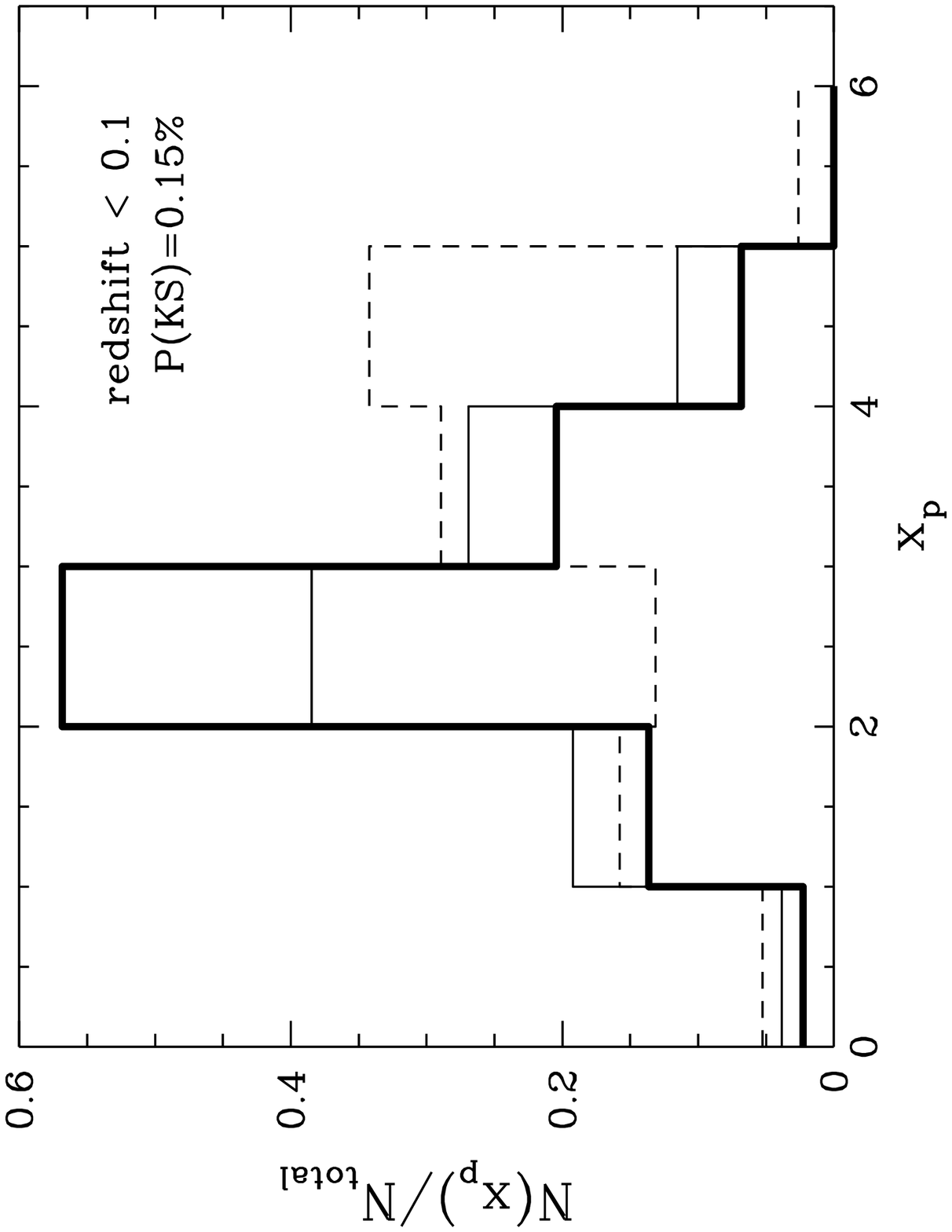}{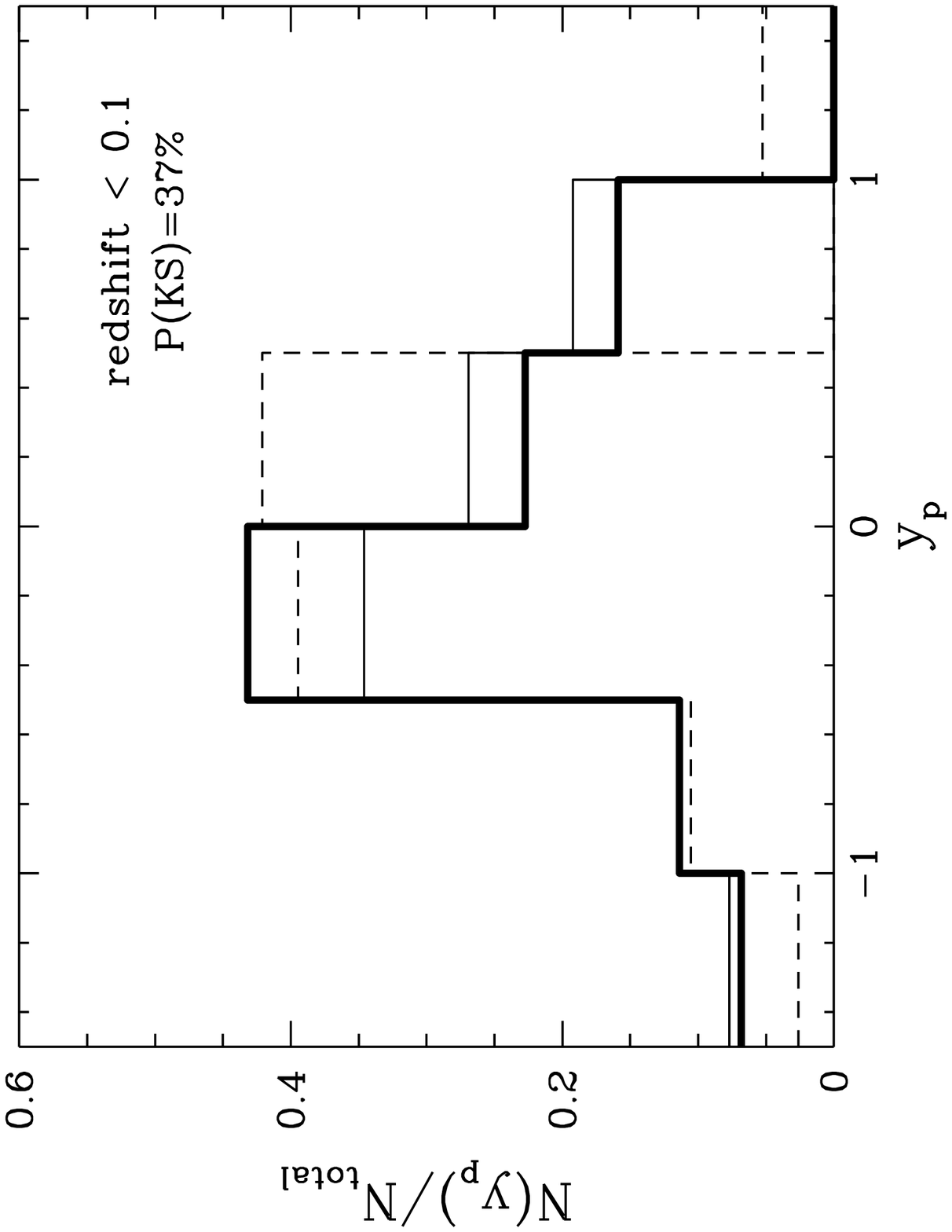}

\raggedright
The distribution of clusters in the $P_3/P_0$ -- $P_2/P_0$ plane are
shown.  Panel (a) shows the number (normalized to the total number
of clusters in the given sample) of clusters with given
$x_p$.  Panel (b) shows the number (normalized to the total) with 
given values of $y_p$.  The histograms are plotted as in Figure
\ref{histo420}
\end{figure}

\begin{figure}
\caption{  \label{histop10} }

\plotone{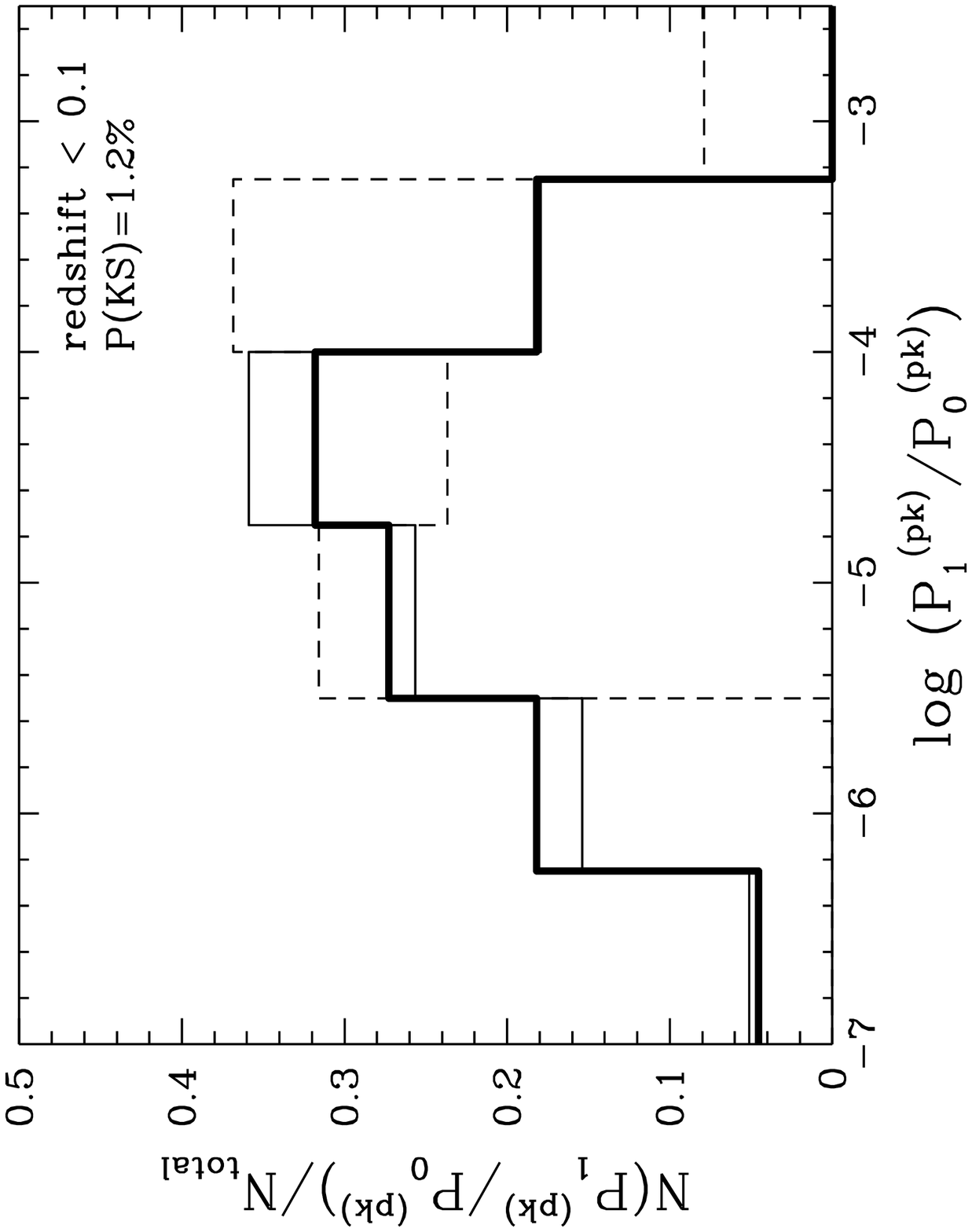}

\raggedright
The number of clusters (normalized to the total number in the sample)
with $P_1/P_0$ in the given range is plotted versus $P_1/P_0$.  
The histograms are plotted as in Figure \ref{histo420}.
\end{figure}

\begin{figure}
\caption{  \label{histo42z} }

\plottwo{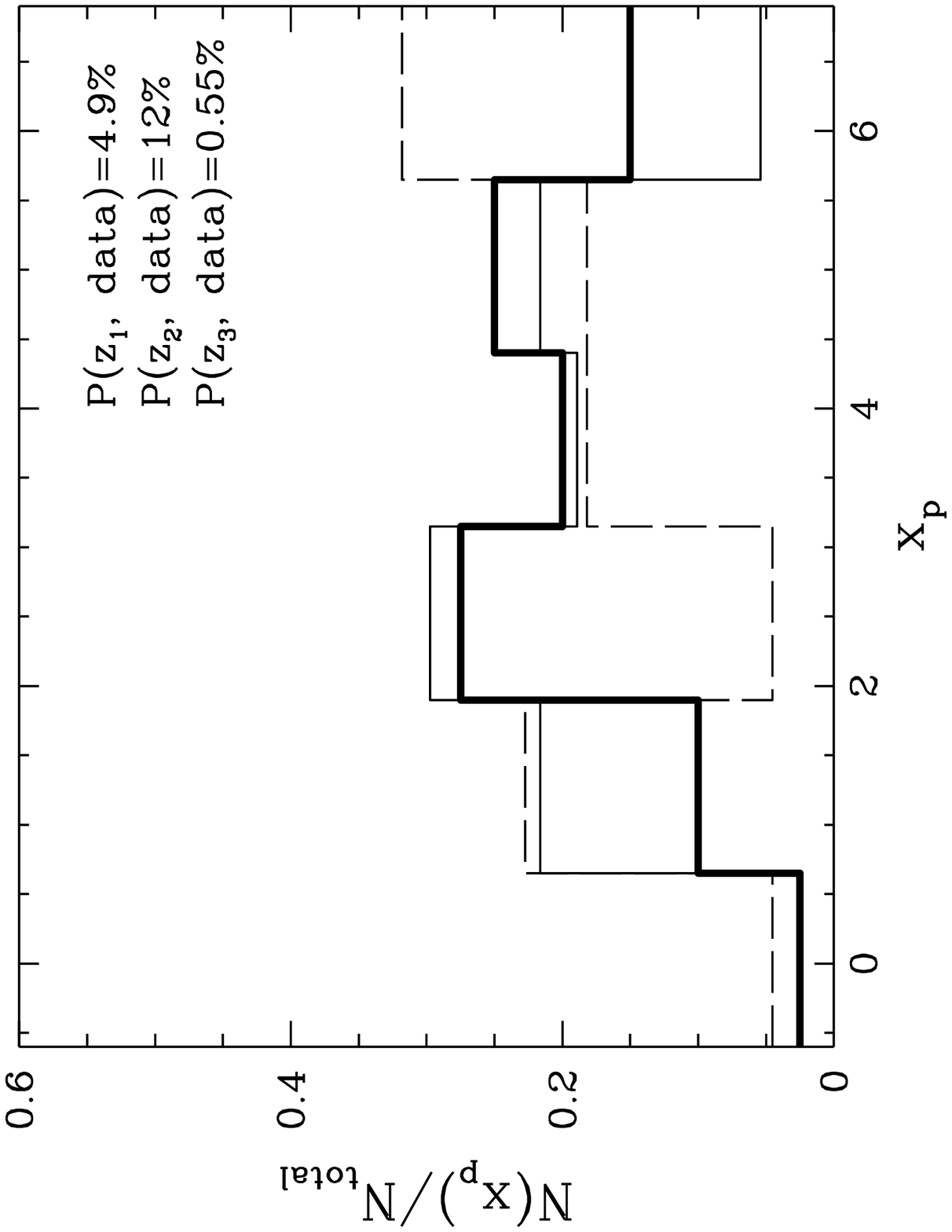}{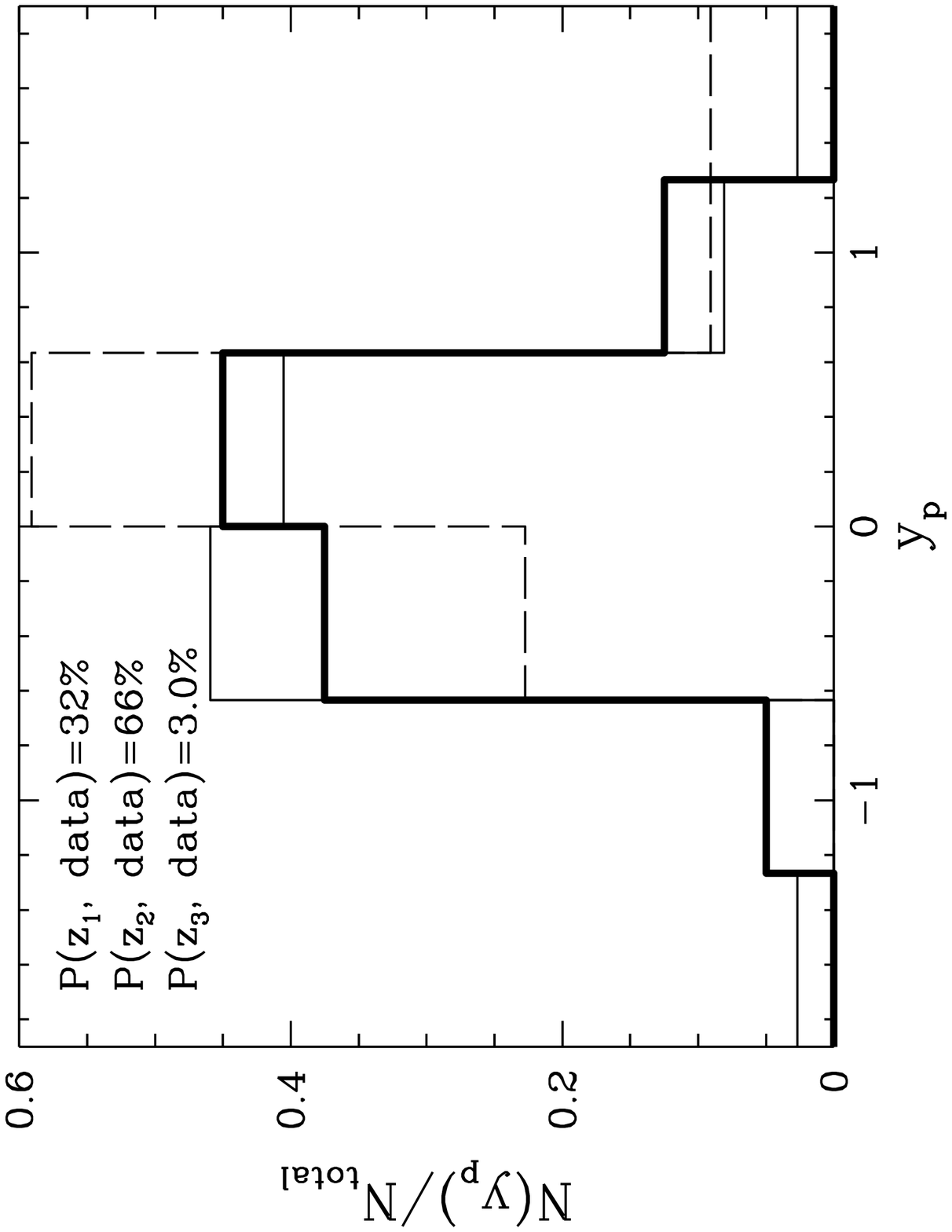}

\raggedright
The distribution of clusters in the $P_4/P_0$ -- $P_2/P_0$ plane are
shown in the $x_p$ (panel [a]) and $y_p$ directions (panel [b]).
The heavy solid histogram gives the distribution for the redshift $z_1$
($z = 0.14$ and 0.23) clusters, the light solid histogram
corresponds to the $z_2$ clusters ($z = 0.35$ and 0.49),
and the dashed histogram corresponds to the $z_3$ clusters ($z = 
0.68$).
\end{figure}

\begin{figure}
\caption{  \label{histo32z} }

\plottwo{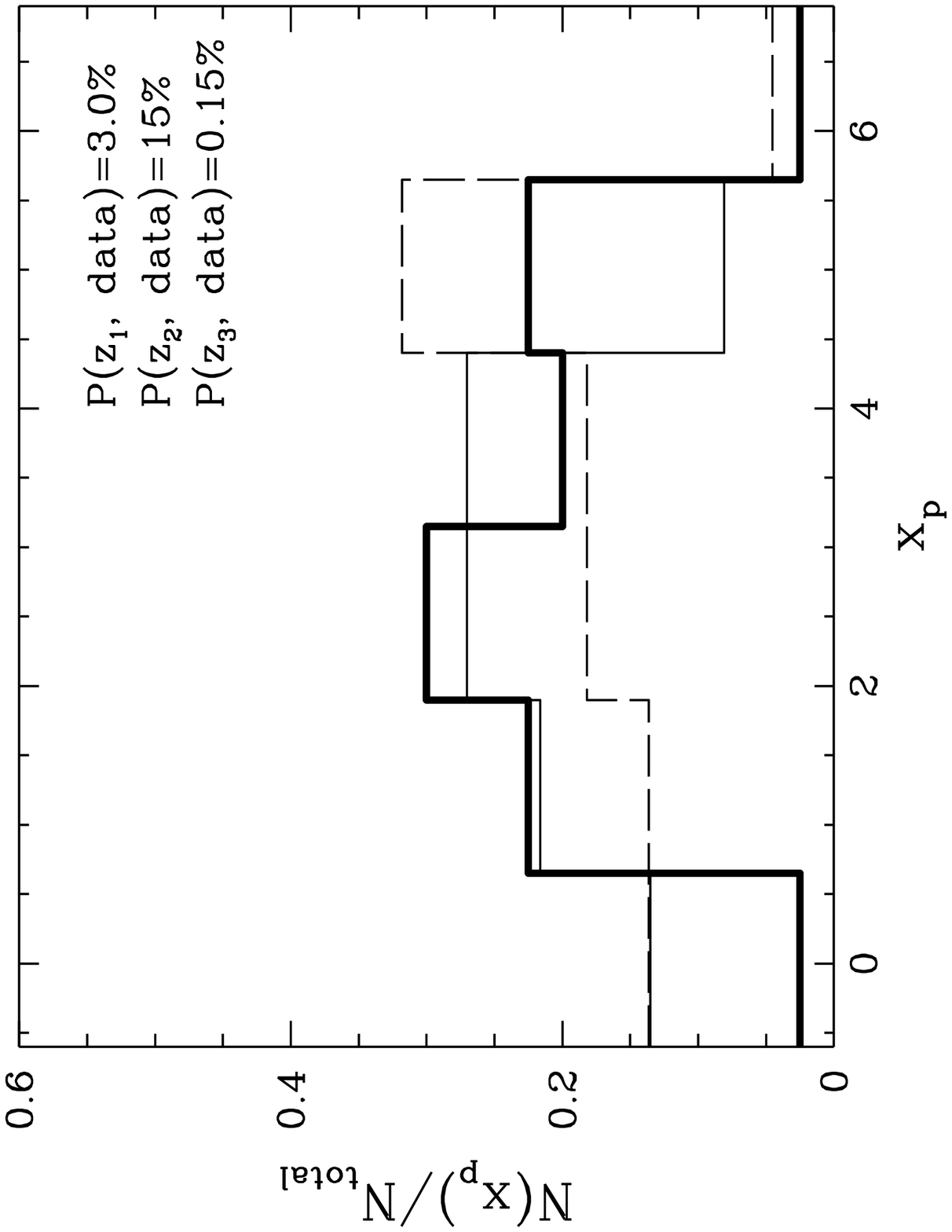}{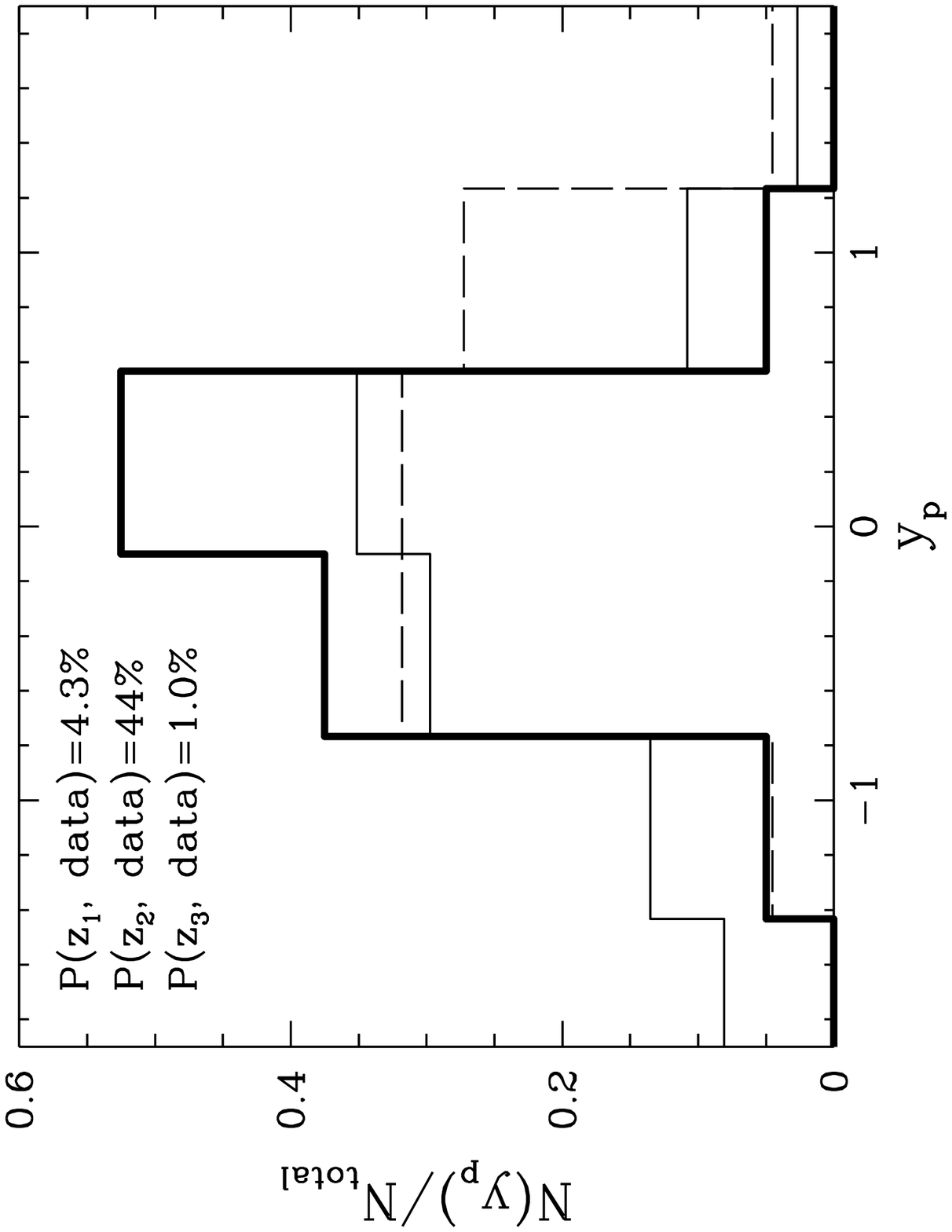}

\raggedright
The distribution of clusters in the $P_3/P_0$ -- $P_2/P_0$ plane are
shown in the $x_p$ (panel [a]) and $y_p$ directions (panel [b]).
Histograms are plotted as in Figure \ref{histo42z}.
\end{figure}

\begin{figure}
\caption{  \label{histop1z} }

\plotone{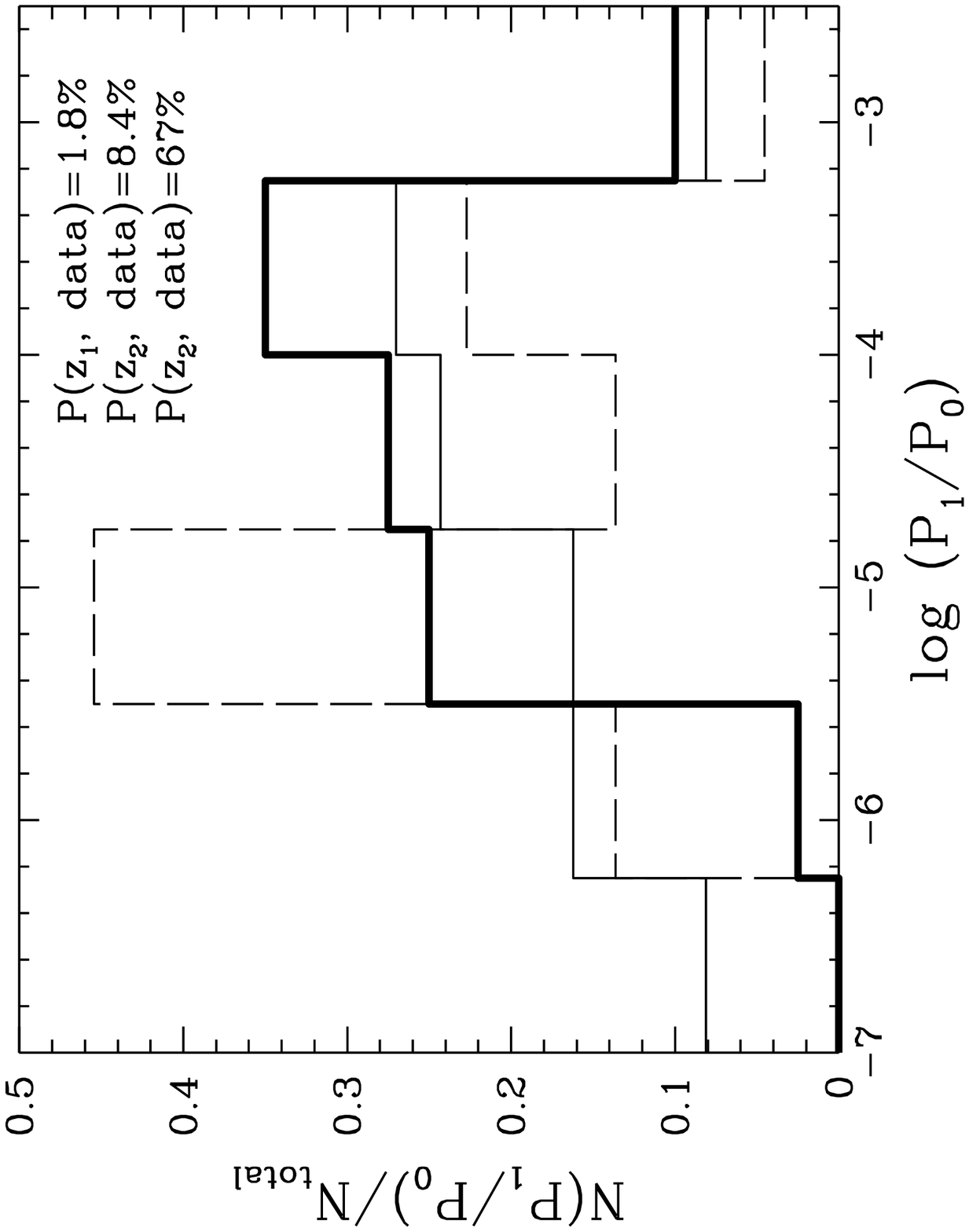}

\raggedright
The number of clusters (normalized to the total number in the sample)
with $P_1/P_0$ in the given range is plotted versus $P_1/P_0$.
Histograms are plotted as in Figure \ref{histo42z}.
\end{figure}

In the $z$ projection the CL1 cluster evolves very nearly along a line
in the $P_4/P_0$--$P_2/P_0$ plane with a slope that agrees well with
that of the best-fit line to the ROSAT data.  We consider simple
parametrized models of the simulated cluster in order to understand
how the cluster adjusts its morphology to follow the
evolutionary track displayed by the ROSAT clusters.  At early times,
the two components of the bimodal cluster are well separated so can be
considered individual clusters for the present purposes.  The surface
brightness profile of each component is fitted to the $\beta$ model
(eq. \ref{beta.mod}) assuming circular isophotes $\epsilon_x=0$.  At
later times ($z\la 0.49$ -- panel (f) of Figure \ref{clus1z})
the merged remnant resembles a rather high ellipticity $\beta$ model
and is consequently fitted with finite $\epsilon_x$.

Table \ref{bet.params} gives the best fit parameters at various
redshifts.  Consider first the bimodal cluster at $z=0.95$ and
0.68.  The top component (the uppermost clump in Figure \ref{clus1z})
evolves little with both $\beta$ and $a_{core}$ retaining steady and
realistic (e.g., Jones \& Forman 1984) values.  The central clump has
a constant $a_{core}$ but a surface brightness profile which becomes
shallower in time.  This simple modification to the naive view that a
bimodal cluster evolves only due to the mutual approach of its two
unchanging components allows the line followed by the cluster in power
ratio space to agree with observations.  Specifically, the path
indicated by the open stars of Figure \ref{powrat42.fit} is modified,
for example, by allowing $\beta$ for one of the clumps to decrease
with time, so that it becomes shallower, in agreement with the
distribution of observed clusters.  An interesting aspect of this
result is that the rate at which $\beta$ changes must be accurately
tuned, in concert with the approach speed of the two clumps, so that
the corresponding path followed in power ratio space is correct.  The
rate of evolution of $\beta$ likely depends primarily on the actions
of self gravity in the given subclump. However, the rate of approach
of the subclumps may be controlled or at least influenced by the large
scale matter distribution, tidal fields, or the underlying
cosmological model.  Thus the slope of the correlation line
potentially gives information both on the rate of gravitational
relaxation and on cosmology, but it will require further simulations
to disentangle the various dependences.

A further consequence of the above parameter evolution is that
simulations of cluster mergers are only realistic if the approach
speed of the two clumps is correctly chosen so that the cluster
satisfies the above evolutionary track constraint.  Subclumps cannot
approach each other at arbitrary speeds, otherwise they will evolve
into a region of power ratio space where no clusters are observed to
reside.  This new restriction is important for many issues which are
addressed by simulations.  For example, simulations based on
cosmologically motivated initial conditions show that cluster gas
relaxes to hydrostatic equilibrium on relatively short time scales
(e.g., Tsai, Katz, \& Bertschinger 1994, NFW).  In contrast, some
simulations concentrating on the study of specific mergers indicate
that relaxation could take much longer (e.g., Roettiger, Burns, \&
Loken 1993; Nakamura et al. 1995).  However, the subclumps in these 
studies are allowed to
merge at rather high speeds (a few times the sound speed).  It is not
clear that the clusters in these simulations would obey the
observational constraints described here and thus whether they
represent realistic cluster simulations.

The evolutionary track also places constraints on the evolution of
single-component clusters.  The simulated cluster at $z=0.49$
and 0.35 are fit to single $\beta$ models.  (At $z=0.49$, two
defined emission peaks remain, however, the separation is very small
$r_{sep}\sim 300 $ kpc so our single cluster assumption is adequate
for the present purposes.) Table \ref{bet.params} shows that the
evolution in this case consists of the surface brightness profile
becoming both shallower and more centrally concentrated.  This
modification to the naive view that single clusters evolve simply by
becoming rounder is again sufficient to ensure that the cluster
follows the required path in power ratio space.  As before,
gravitational relaxation likely leads to the required parameter
evolution of the cluster, however, external tidal fields could also
play a role.

\section{Comparison of Dark Matter and Gas\label{dark.gas}}

\begin{figure}
\caption{  \label{darkmatpowrat}}

\plottwo{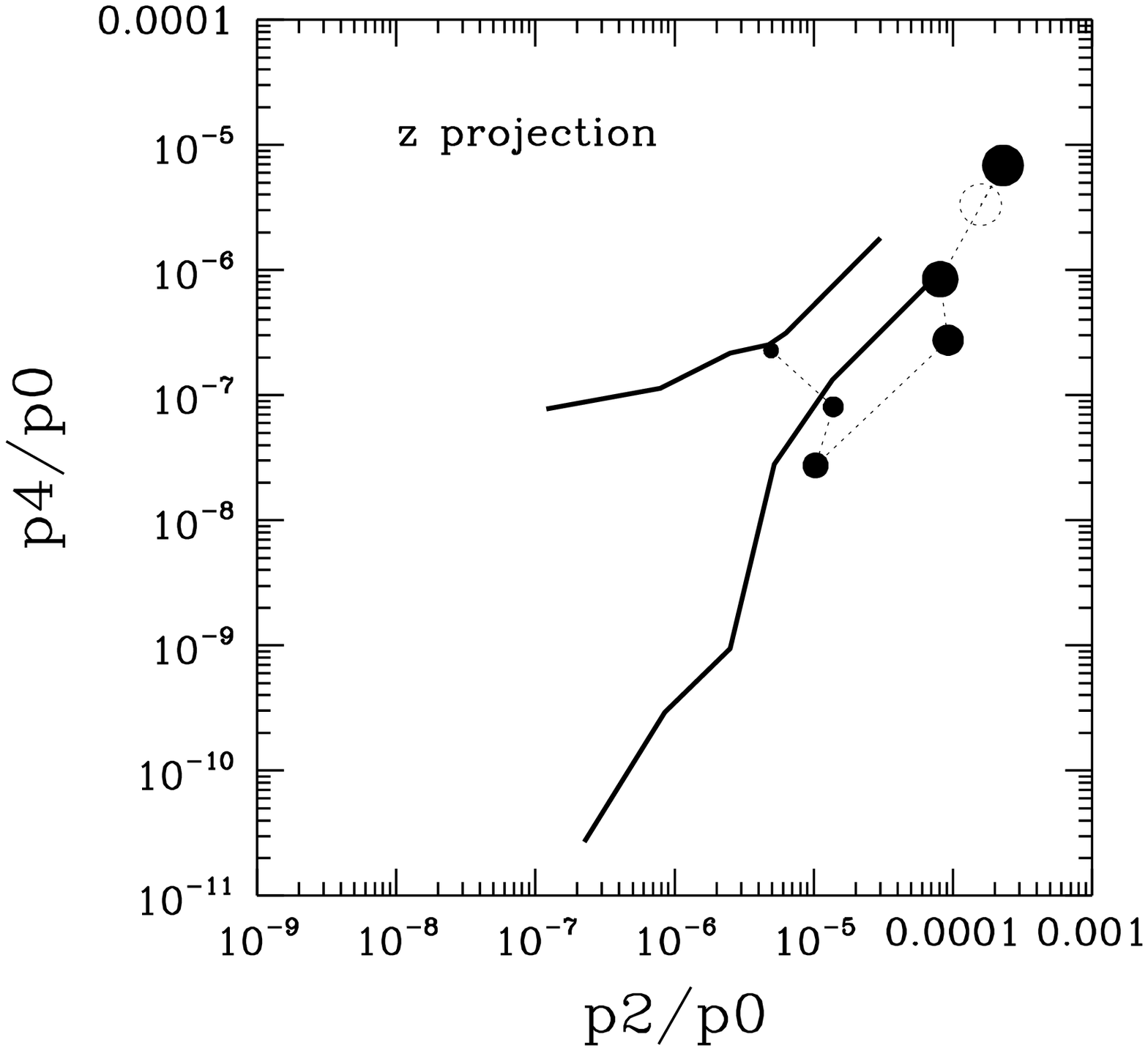}{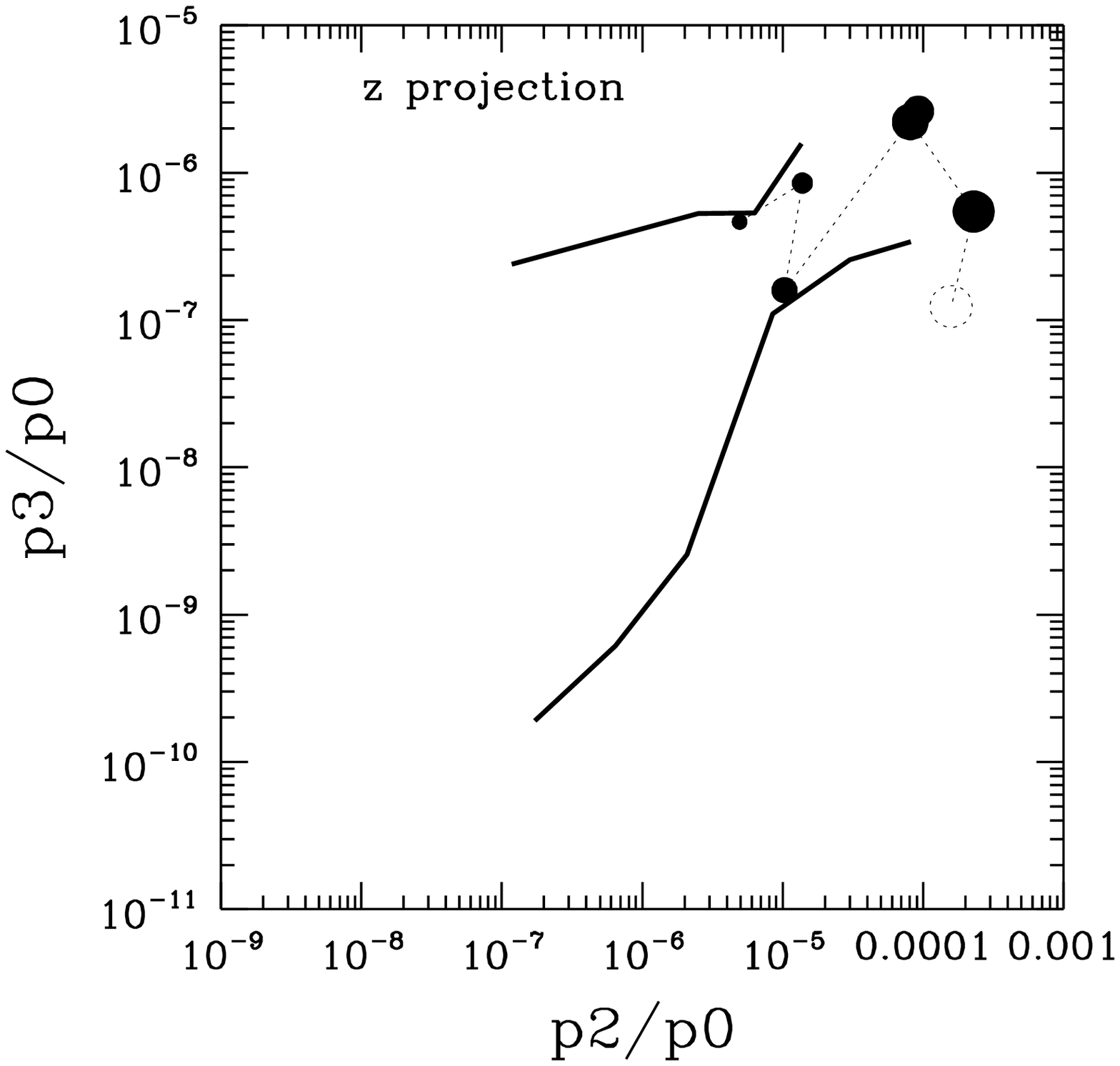}

\raggedright
Power ratios computed for the projected dark matter distribution along
the $z$ axis for the largest clusters CL1.  The ratios are plotted as in
Figure \ref{powrat42}, where (a) shows $P_4/P_0$ versus $P_2/P_0$ and (b)
shows $P_3/P_0$ versus $P_2/P_0$.  An aperture of 1 Mpc is assumed.
\end{figure}

An important assumption underlying our analysis is that the X-ray
emission traces the underlying mass sufficiently well to allow the
power ratios to quantify the evolutionary state of the underlying
mass. Although we expect the distributions to be different in detail
even in rather virialized clusters (e.g, Buote \& Canizares 1996 and
references therein), a simple qualitative examination of the simulated
clusters shows that the gross morphological characteristics of the
dark matter and of the gas are similar (e.g., Buote \& Tsai 1995b;
NFW).  The power ratios allow this connection to be quantified in
detail.  As an example, Figure \ref{darkmatpowrat} shows the power
ratios computed on a 1 Mpc aperture for the dark matter distribution
of the largest cluster CL1.  Specifically, $\Sigma \equiv\int
\rho_{DM} dz$ is used in eqs. (\ref{eqn.mom1}) and (\ref{eqn.mom2})
where $\rho_{DM}$ is the dark matter mass density.  The expansion of
eq. (\ref{eqn.multipole}) in this case will give exactly the projected
potential.

The cluster dark matter is distributed much like the gas shown in
Figures \ref{clus1x}, \ref{clus1y}, and \ref{clus1z} (see also NFW).
Concentrations of dark matter coincide with concentrations of gas and
the correspondence is retained in the subsequent mergers and evolution
of the clumps.  However, after mergers, the distributions of the gas
and dark matter differ in that the dark matter retains a larger
quadrapole moment than does the gas.  In fact, gas distributed in
hydrostatic equilibrium under the influence of an ellipsoidal dark
matter halo is rounder than the underlying mass distribution (Buote \&
Canizares 1996; Buote \& Tsai 1995b).  Thus an exact numerical
correspondence between the power ratios for the gas and the dark
matter is not expected.  The dark matter power ratios will in general
be larger both because it dominates the mass, hence is less ``round''
than the gas, and because it evolves less rapidly since
non--axisymmetric structures are not erased as readily as in the
dissipational gas.  What is important, however, is that the gas and
the dark matter give a consistent picture of the state of the cluster.

The figures show that the dark matter power ratios evolve generally
like those of the gas, although the values extend over a smaller
region of the space, as we expect.  At early times, the two clumps
which form the nascent cluster are widely separated.  As they approach
and merge, the power ratios move down and to the left, although the
values remain large compared to that of the gas.  However, the notion
of the evolutionary track is clearly valid for the dark matter and the
dynamical states as classified by the dark matter and gas are
consistent.  Note, however, that the track for the dark matter does
not seem to coincide with that of the gas.  For example, the track in
the $P_4/P_0$--$P_2/P_0$ correlation falls below that of the gas at
some redshifts.  There is no {\it a priori} reason that the two tracks
should be exactly the same; rather this serves to illustrate that the
different aspects of the evolution of the dark matter and gas may be
usefully quantified and analyzed by the power ratios.  Detailed
statistical analyses should provide interesting information on the
relative behaviors of dark matter and gas as they evolve
simultaneously in the CDM model which has, among others, implications
for dynamical estimates of the mass distributions of galaxy clusters
using X-rays. These issues are beyond the scope of the present paper
and will be considered elsewhere.

\section{Discussion and Conclusions\label{disscon}}

With the aid of hydrodynamical simulations of cluster formation, we
have further demonstrated the utility of the power ratios for
quantifying the morphology and evolution of galaxy clusters and
have shown their promise for providing new constraints on
cosmology.  These statistics are dynamically motivated: each power
ratio measures the square of a higher order moment of the projected
potential relative to the monopole term and hence quantifies the
amount of higher order, non--axisymmetric (for odd moments) structure
present in the projected potential.  The power ratios were computed
elsewhere for a sample of low redshift ($z\lesssim 0.2$)
clusters observed by ROSAT (BTb) and displayed interesting
correlations. The observed clusters occupy only a restricted region of
the space defined by the three lowest order power ratios (log
$P_2/P_0$, log $P_3/P_0$, log $P_4/P_0$), extending in a thick
straight filament outward from the origin.  This filament, or any of
the correlation lines resulting from projection of the filament onto
the coordinate planes, was interpreted in the context of bottom--up
cluster formation scenarios as the evolutionary track followed by
clusters as they form out of the background distribution (moving onto
the track from above), relax (moving down along the track towards the
origin), or merge with a clump previously outside the given aperture
(moving outward from the origin along the track).

The cluster simulations confirm the above picture of the evolutionary
track and show that clusters formed in the standard $\Omega=1$ CDM
cosmogony evolve along the same track as given by the data.  In
addition, a combination of the simulations and simplified parametric
cluster models allow an intuitive understanding of cluster evolution.
A bimodal cluster, for example, would evolve by the mutual approach of
the two components.  A simple model consisting of two unchanging
components (taken to be $\beta$ models) approaching each other follows
a straight line in power ratio space that mimics the qualitative
aspects of the evolutionary track, but has a steeper slope.  The
simulations show that if in addition the components evolve
individually by self gravity giving smaller core radii or shallower
mass distributions, the slope will agree with that of the observed
track.  Since the evolution of the components is likely driven mainly
by self gravity, the rate of approach of the two clumps must be finely
tuned so as to ensure that the cluster moves in an allowed
(i.e. probable) region of power ratio space.  This approach speed may
depend upon the assumed cosmology. Hence the slope of the evolutionary
track may constrain the underlying cosmology.  A large suite of
simulations with many different cosmogonies must be examined to
disentangle the various dependences.

BTb asserted that the distribution of ROSAT clusters along the
evolutionary track is a measure of the balance of the formation and
relaxation rates of clusters. Guided by the success of the
evolutionary track interpretation shown by the simulated clusters, in
this paper we explicitly defined these rates in terms of the
evolutionary track. We considered the ``relaxation rate'' to be the
hypothetical rate at which clusters would move down the evolutionary
track in isolation; i.e. when undisturbed by mass accreting from
outside the specified aperture.  We defined the ``formation rate'' to
consist of a ``birth rate'' and ``merger rate''.  The birth rate is
the rate at which clusters first appear on the upper parts of the
track immediately after forming out of the background matter while the
``merger rate'' is the rate at which a mature cluster jumps from a
lower position to a higher position on the track due to the accretion
of mass from outside the specified aperture. Provided these rates have
physical meanings similar to conventional definitions, then this would
suggest that to probe $\Omega$ effectively we should consider the
location of clusters in power-ratio space relative to a set of rotated
coordinates where one axis lies along the correlation line.

We compared the distribution of simulated clusters along and
perpendicular to the evolutionary track with the observed
distributions of ROSAT clusters.  It was therefore useful to consider
a rotated coordinate system $(x_p, y_p)$ where $x_p$ coincided with
the evolutionary track and $y_p$ was taken to be perpendicular to it.
The simulated cluster sample at low redshift contained too many large
$x_p$ clusters relative to the observations.  Provided that our
assumptions regarding the construction of the simulated cluster sample
are not grossly inaccurate (e.g., that each simulated cluster
represents the ``average'' cluster of that size) then the $\Omega=1$
cosmogony produces too many clusters with significant substructure.
This latter conclusion is supported by ongoing work on dissipationless
simulations where the simulated clusters sample is much more
statistically significant (Buote \& Xu 1996). Note these conclusions
differ from previous studies (e.g., Mohr et al. 1995; Jing et
al. 1995). 

Since the distribution in the $x_p$ direction results from the
combined actions of formation and relaxation, either the formation
rate of clusters in the simulations is too high, or the relaxation
rate is too low.  Since relaxation is treated self-consistently, it is
likely that the formation rate with $\Omega=1$ is too high.
$\Omega<1$ cosmogonies are thus preferred, given the arguments of
Richstone et al. (1992), and agrees with results from recent cluster
dynamical studies (e.g., Carlberg et al. 1995), over-density of
baryons in clusters (e.g., White et al. 1993; Buote \& Canizares
1996), and galactic disk arguments (Toth \& Ostriker 1992), but does
not agree with analytic treatments of cluster collapse (e.g.,
Richstone et al. 1992; Lacey \& Cole 1993).  Part of this latter
disparity may result from the use of qualitative estimates of the
fraction of clusters possessing substructure in place of quantitative
statistics employed in the present study and because of the reliance
on linear theory results.  The application of power ratios to analytic
treatments of cluster collapse may clarify the reasons for the above
disagreement as well as elaborating on the $\Omega$ dependence of
cluster evolution.  These arguments will be presented elsewhere.  It
will in any case also be useful to confirm the present results with
more statistically complete samples of simulated clusters.  This
necessarily implies that we consider purely dissipationless dark
matter simulations where the computational demands are tractable
(Buote \& Xu 1996).

The distributions of the simulated and observed ROSAT clusters agree
in the $y_p$ direction; i.e. perpendicular to the evolutionary track.
For a given point on the evolutionary track a cluster lying
perpendicularly to the left (i.e. positive $y_p$) of the track
possesses a larger amount of smaller scale structures (i.e. excess
$P_3/P_0$ or $P_4/P_0$) but a less pronounced aggregate, larger scale
structure (i.e. less $P_2/P_0$); the opposite is true for clusters
lying perpendicularly to the right (i.e. negative $y_p$) of the track.
Intuitively the agreement of the ROSAT and simulated cluster
distributions in $y_p$ implies that for a particular dynamical age
(i.e. at a particular position along the evolutionary track) simulated
clusters in the $\Omega=1$ simulation tend to ``look'' like real
clusters. Unlike the cluster distributions in the $x_p$ direction, the
physical implications of this are unclear at present.  One possibility
is that the shape of the power spectrum of initial perturbations has
been correctly chosen for the simulations so as to give the observed
distribution of structure on various scales inside the cluster.
Alternatively, the observed distribution may be simply a general
feature of gravitational relaxation.  We cannot determine which of the
two possibilities is more likely without recourse to simulations with
a variety of power spectra (Buote \& Xu 1996). However, the study of
this interesting aspect of cluster evolution has become possible with
the power ratios.

We examined how cluster evolution in the CDM model changes with time
and suggested corresponding observational tests of the model.  We
found that the distributions of simulated clusters in both the $x_p$
and $y_p$ directions are essentially constant from the present up
until a redshift of $\sim 0.6$.  We caution that this result may be
uncertain due both to the small number of simulated clusters and
because the selection criterion for the clusters at high redshifts was
not entirely consistent with that of the observed sample of clusters.
However, if the above results hold, then, when coupled with the
assumption that the relaxation rate is nearly constant with redshift,
the formation rate (i.e. birth rate + merger rate, as above), is
constant over the indicated period of time.

At $z \gtrsim 0.6$ the clusters in the simulations primarily consist
of single unrelaxed concentrations of mass.  At later times, however,
cluster evolution is dominated by the merger of nearly equal-sized
clumps, as previously indicated by other studies (e.g. Lacey \& Cole
1993).  Then at early times clusters consist of single lumps of gas
and dark matter forming out of the background distribution.  At $z\sim
0.6$, the merger of the nascent clumps becomes the dominant means of
cluster growth.  The rate of mergers appears basically constant from
that time onwards to the present -- a result unanticipated by analytic
studies of cluster evolution (e.g., Richstone et al. 1992).  Whether
the constancy in the merger rate (as defined by the power ratios) is a
general feature of hierarchical clustering models, or if the redshift
at which mergers begin to dominate cluster evolution changes with
model parameters, awaits further study. However, the above results
suggest that the era around $z \sim 0.6$ may have observationally
interesting properties which can be considered with the expected high
quality data from AXAF.
\acknowledgements
We are grateful to J. F. Navarro for graciously allowing us to use his
cluster simulations as the basis of this paper. DAB acknowledges the
hospitality of CITA during the final stages of this work.

\vfill 
\eject
% ********************* TABLES ******************************************
\begin{table}
\caption{Cluster Parameters\label{bet.params}}
\begin{tabular}{ccccccccccccc}  \tableline\tableline

$z$ & Clump & $r_{sep}$ (kpc) & $\beta$ & $a_{core}$ (kpc) \\
\tableline
0.95 & top     & 1230. & 0.852 & 92.1 \\
0.95 & central & 1230. & 0.938 & 110. \\
0.68 & top     & 1050. & 0.844 & 94.8 \\
0.68 & central & 1050. & 0.839 & 104. \\
0.49 & central &  --   & 1.34 & 629. \\
0.35 & central &  --   & 0.864 & 325. \\
\tableline

\tablecomments{ The best fit $\beta$ model parameters are given
for various clusters and cluster components.  The second column
indicates either the top or central component of the bimodal cluster
in Figure \ref{clus1z} or the central cluster after $z=0.49$.  The
third column gives the separation of the cluster components.
 }

\end{tabular}

\end{table}
%
%****************************References*******************************

%*********************************Figures*******************************
\clearpage

\clearpage
\begin{references}
\reference Bartelmann, M.,Ehlers, J., \& Schneider, P. 1993, \aap, 280, 
351

\reference Briel, U. G., \& Henry, J. P. 1993, \aap, 278, 379

\reference Buote, D. A., \& Canizares, C. R. 1996, \apj, 457, 565

\reference Buote, D. A., \& Tsai, J. C. 1995a, \apj, 452, 522 (BTa)

\reference Buote, D. A., \& Tsai, J. C. 1995b, \apj, 439, 29

\reference Buote, D. A., \& Tsai, J. C. 1996, \apj, 458, 27 (BTb)

\reference Buote, D. A., \& Xu, G. H. 1996, in preparation

\reference Castander, F. J., Ellis, R. S., Frenk, C. S., Dressler, A., \&
Gunn, J. E. 1994, \apj, 424, L79

\reference Carlberg, R., Yee, H. K. C., Ellingson, E., Abraham, R., Gravel, P.,
Morris, S., \& Pritchet, C. J. 1995, preprint

\reference Cavaliere, A., \& Fusco--Femiano, R. 1976, \aap, 49, 137

\reference Donahue, M.,\& Stocke, J. T. 1995, \apj, 449, 554

\reference Dressler, A. \& Shectman, S. A. 1988, \aj, 94, 985

\reference Edge, A. C., Stewart, G. C., Fabian, A. C., \& Arnaud,
K. A. 1990, \mnras, 245, 559

\reference Evrard, A. E., Mohr, J. J., Fabricant, D. G., \& Geller, 
M. J. 1993, \apj, 419, 9

\reference Geller, M. J., \& Beers, T. C. 1982, \pasp, 94, 421

\reference Jing, Y. P., Mo, H. J., B{$\ddot{\rm o}$}rner, G., \& Fang,
L. Z. 1995, \mnras, in press

\reference Jones, C., \& Forman, W. 1984, \apj, 276, 38

\reference Kaiser, N., \& Squires, G. 1993, \apj, 404, 441

\reference Kauffmann, G., \& White, S. D. M. 1993, \mnras, 261, 921

\reference Lacey, C., \& Cole, S. 1993, \mnras, 262, 627

\reference Mohr, J. J., Fabricant, D. G., \& Geller, M. J. 1993, \apj,
413, 492

\reference Mohr, J. J., Evrard, A. E., Fabricant, D. G., \& Geller, 
M. J. 1995, \apj, 447, 8

\reference Nakamura, F. E., Hattori, M., \& Mineshige, S. 1995, \aap, in
press

\reference Navarro, J. F., Frenk, C. S., \& White, S. D. M. 1995, \mnras,
275, 720 (NFW)

\reference Pfeffermann, E. et al. 1987, Proc. SPIE, 733, 519

\reference Roettiger, Burns, \& Loken 1993, \apj, 407, L53

\reference Richstone, D. O., Loeb, A., \& Turner, E. L. 1992, \apj,
393, 477

\reference Squires, G., Kaiser, N., Babul, A., Fahlman, G., Woods, D.,
Neumann, D. M., \& B{$\ddot{\rm o}$}hringer, H. 1995, \apj, submitted

\reference Toth, G., \& Ostriker, J. P. 1992, \apj, 389, 5

\reference Tsai, J. C., Katz, N., \& Bertschinger, E. 1994, \apj, 423, 553

\reference West, M. J., \& Bothun, G. D. 1990, \apj, 350, 36

\reference White, S. D. M., Navarro, J. F., Evrard, A. E., \& Frenk, C. S.
1993, Nature, 366, 429

\end{references}
\end{document}